\documentclass[12pt,draftclsnofoot,letterpaper,onecolumn,romanappendices]{IEEEtran}
\usepackage{cite}
\usepackage{epsfig}
\usepackage{amsthm}
\usepackage{amsmath,amssymb,amsfonts,amstext,amsbsy,amsopn,dsfont}
\usepackage{cases}
\usepackage{bigints}
\usepackage{color}
\usepackage{sublabel}
\usepackage{array}

\newtheorem{proposition}{\textbf{Proposition}}
\newtheorem{lemma}{\textbf{Lemma}}

\newtheorem{algorithm}{\textbf{Algorithm}}

\newcommand{\defn}{\triangleq}
\newcommand{\dif}{\textmd{d}}

\newcommand{\ul}{\mathsf{ul}}
\newcommand{\dl}{\mathsf{dl}}

\begin{document}

\title{\LARGE Full-Duplex Heterogeneous Networks with Decoupled User Association: Rate Analysis and Traffic Scheduling}

\author{Chun-Hung Liu and Heng-Ming Hu\\
	\thanks{C.-H. Liu is with the Department of Electrical and Computer Engineering at Mississippi State University. H.-M. Hu is with the Department of Electrical and Computer Engineering at National Chiao Tung University, Hsinchu 30010, Taiwan. Part of this paper was presented in IEEE International Conference on Communications, May 2018 \cite{CHLHMH18,HMHCHL18}. The contact author is Dr. Liu  (e-mail: chliu@ece.msstate.edu). }}

\maketitle

\begin{abstract}
Full-duplex (FD) transmission in a point-to-point  (P2P) link, wherein bidirectional traffic flows simultaneously share the same spectrum, has the capability of doubling the link rate by completely removing self-interferences. However, the rate performance of an FD heterogeneous network (HetNet) is not as clear as that of an FD P2P link due to the co-channel interferences induced by complex FD and half-duplex (HD) transmission behaviors in the HetNet. To thoroughly investigate the achievable link rate performances of users and base stations (BSs) in a HetNet with decoupled user association, a hybrid-duplex approach is proposed to model a HetNet in which all BSs and users can perform HD or FD transmission depending on their traffic patterns. We first characterize the decoupled rate-optimal user association scheme and use it to define and evaluate the downlink and uplink rates in the HetNet. The tight lower bounds on the link rates of the FD users and BSs are found in a neat form that characterizes general channel fading, imperfect self-interference cancellation and the intensities of users and BSs. These bounds outline the rate regions of the FD users that inspire us to propose the opportunistic FD scheduling algorithms that maximize the sum rate of each bidirectional traffic and stabilize each of the queues in the HetNet.    
\end{abstract} 


\section{Introduction}
As more and more smart handsets and mobile devices are widely adopted,  a tremendous data traffic demand in the next generation (5G) cellular networks is surely foreseen. How to make 5G cellular networks carry such a huge traffic demand becomes a very thorny problem that needs to be dealt with immediately and carefully.  Essentially, a 5G cellular network is a heterogeneous network (HetNet) since it will consist of different kinds of base stations (BSs) using different radio-access technologies \cite{AndrewsEtal14}. Although such a HetNet is able to achieve large network throughput by densely deploying BSs, its throughput performance is eventually dominated by limited available spectrum resources \cite{MAARNS16}. Full-duplex (FD) transmission, wherein a transceiver can simultaneously transmit and receive information over the same spectrum, seems to be an effective tonic to alleviate the spectrum crunch crisis in HetNets if its performance hurdles due to intrinsic self-interference and co-channel interference can be cleverly overcome\cite{CLICRSHXGLZP14,DKHLDH15,SGPLSSPRADRYEB15}. To successfully apply FD transmission in a HetNet, we thus need to largely suppress self-interference as well as co-channel interference so that the considerable rate gain offered by FD can be exploited as much as possible. 

To delve whether or not FD transmission essentially benefits the rate in a HetNet, in this paper we aim at thoroughly investigating the downlink and uplink rate performances of an FD HetNet  in which each base station (BS) and its serving user can perform FD transmission whenever there exists bidirectional traffic between them. We consider decoupled user association in the HetNet that allows users associate with different BSs in downlink and uplink since it induces the flexibility in user association that may improve users'  link performance\cite{FBJAHEMDSPPP16}. Also, to make our rate analysis more realistic, imperfect self-interference cancellation and co-channel interference are both considered in the total interference model. In particular, the co-channel interference model considers the void cell issue that is induced by user-centric user association and results in the phenomenon that some BSs in a HetNet are not associated with any users\cite{CHLLCW16}. Although such a co-channel interference with void cell modeling is a more correct and accurate model, it is seldom studied in the literature\cite{CHLLCW1502}.

\subsection{Prior Related Work and Motivations}
Some of prior works on the comprehensive rate analysis in FD cellular networks are mainly built based on a single cell network model. Reference \cite{JSHJSBSJJSS14}, for example, focused on how to do cooperative communication in a single cell network to achieve full spatial diversity and reference \cite{DNLNTPPMLA14} studied the deterministic spectrum efficiency in a single-cell FD network. Reference \cite{DGWNASCASMSA17} studied how to select a suitable antenna set for maximizing the average transmission rate, reference \cite{1IRHDHEMSA7} looked into the problem of maximizing average link rates by managing interference through user association, scheduling, power control and spectrum allocation, and reference \cite{PXJJCZXWXZ17} studied how to optimize the transmit power of users by simply considering a multi-pair two-way FD relay network with a large-scale antenna array. In  reference \cite{TRSWRW11}, a fundamental trade-off problem of using either FD mode or HD mode in a relay link was investigated in a single relay network and an opportunistic switching scheme between FD and HD modes was proposed. In reference \cite{CLBXQJYYGY18}, the achievable rate of a two-way FD relay system with multiple users was studied. Since these prior works were merely developed in a single-cell FD network with a fixed number of BSs or a single BS/relay so that how their analytical results are affected by multi-tier interferences and network heterogeneity cannot be clearly perceived. 

By comparing with the aforementioned previous works, there are indeed some prior recent works that consider a large-scale FD cellular network model (typically see \cite{JLTQSQ15,SHPLSSP17,YSDWKNZDRS17,SDJPCGC17,XQHZXYBJYTHWLSFM17,YLPFALLL17,IAMK17,PAACCTR17}). In reference \cite{JLTQSQ15}, for instance, the network throughput was studied in a HetNet consisting of multi-tier HD and FD access points, but the average link rates were not studied and they are assumed as a constant to characterize the network throughput, which may not be an accurate and proper approach to characterizing the network throughput in that the average link rates dominated by the interferences in an FD network are hardly a constant. Reference  \cite{SHPLSSP17} studied a joint uplink and downlink scheduling problem in a single-tier multi-cell network and it aimed to maximize the network throughput by optimally doing user scheduling and power allocation in a distributed fashion. These works do not consider how different decoupled/coupled user association schemes and FD traffic scheduling schemes influence their rate analysis even though their network models characterize some generality and complexity of FD HetNets. 

The decoupled user association problem has recently attracted some attentions and been studied in a few recent works. Reference \cite{SSXZJGA15} considered a decoupled uplink-downlink biased cell association to analyze the rate coverage in a HetNet with load balancing and power control. Reference \cite{MBYWBC17} studied how to improve the coverage probability in a two-tier HetNet with multi-antenna BSs and decoupled user association. In reference \cite{SSHTEH17}, the decoupled user association problem was formulated as a matching game in a two-tier FD cellular network with some performance constraints and it was applied to solve the rate maximization problem. Reference \cite{AHSEH17} considered a multi-tier in-band FD network with decoupled user association and studied an optimization problem that aims at maximizing the mean rate utility. These prior works do not clarify if using FD transmission \textit{all the time} in the network really benefits the total network throughput. 


\subsection{Contributions}
Although FD transmission is principally able to bring considerable rate gain for a P2P link, its fundamental link rate limits are not completely studied yet in a HetNet with decoupled user association. In this paper, our main goal is to provide a clear and good picture on when and how to use FD transmission in a HetNet with decoupled user association so as to benefit the link rates of users. The contributions of achieving this main goal are summarized in the following: 
\begin{itemize}
	\item We consider an FD HetNet model in which decoupled user association is allowed and the total interference model characterizes the impacts from imperfect self-interference cancellation and co-channel interference with void cell modeling. Using this model to perform our link rate analysis can lead to more general and accurate analytical outcomes close to the authentic fundamental limits if it is compared with the FD network models in the literature.  
	\item We provide a novel and generalized analytical approach to characterizing and analyzing the downlink and uplink rates. With the aid of the integral identity of the Shannon transformation found in our previous work \cite{CHLHCT17}, we successfully derive the tight lower bounds on the downlink and uplink rates of an FD link between a user and its associated BSs when the decoupled generalized  user association scheme is adopted in the HetNet. The salient characteristic of the derived bounds is their generality in that they are derived without assuming any specific channel gain models, user association schemes, and FD traffic patterns between users and their associated BSs.  
	\item We find the decoupled rate optimal user association scheme and then use it to derive the tight lower bounds on the maximum uplink and downlink rates of an FD link. These bounds are numerically verified their tightness and accuracy. Also, the tight bounds on the link rate help us to show that using FD transmission all the time in a HetNet may not benefit the sum of the downlink and uplink rates of users, which motivates the ideas of finding the rate region of the uplink and downlink rates of an FD link. 
	\item We use the derived tight bounds on the uplink and downlink rates of an FD link to characterize the rate regions of an FD link with different downlink and uplink traffic patterns and the rate region indicates the maximum rate region of an FD link that can be achieved by  properly and opportunistically adopting HD and  FD transmissions.  According to the observations drawn from the rate regions, we propose two FD opportunistic scheduling algorithms for downlink and uplink to achieve the maximum rate region and show that all downlink and uplink queues in the HetNet can be stabilized by these two scheduling algorithms. Finally, we provide some numerical results to verify that the proposed FD opportunistic scheduling algorithms indeed achieve the maximum rate of an FD link
\end{itemize}
  Furthermore, we also provide some numerical results to validate the correctness and accuracy of our analytical findings and demonstrate the found rate regions and their achievability through our proposed two scheduling algorithms. 


\section{Full-Duplex Network Model and Preliminaries}\label{Sec:NetworkModel}
Consider an interference-limited HetNet on $\mathbb{R}^2$ in which all users form an independent Poisson point process (PPP) $\mathcal{U}$ of intensity $\mu$ given by
\begin{align}
\mathcal{U}\defn \{U_j\in\mathbb{R}^2: j\in\mathbb{N}_+\},
\end{align}
where $U_j$ denotes user $j$ and its location. Each user can perform either FD or HD transmission mode -- it performs the FD mode if it wants to simultaneously exchange information with its associated BS; otherwise it performs the HD mode to merely receive or transmit data\footnote{Throughout this paper, the users/BSs that perform the FD mode are called ``FD users/BSs" and the other users/BSs who perform the HD mode are called the ``HD users/BSs''. }. This HetNet is comprised of $M$ different tiers of base stations (BSs) and the BSs in each tier are of the same type and performance. The first tier consists of macrocell BSs, whereas the rest of $M-1$ tiers consist of small cell BSs, e.g., picocell, femtocell BSs, etc. Specifically, the BSs in the $m$th tier form an independent homogeneous PPP $\mathcal{X}_m$ of intensity $\lambda_m$ given by
\begin{align}
\mathcal{X}_m\defn\{ X_{m,i}\in\mathbb{R}^2: i\in\mathbb{N}_+\},\, m\in\mathcal{M}\defn\{1,2,\ldots,M\},
\end{align}
where $X_{m,i}$ denotes BS $i$ in the $m$th tier and its location. Every BS can also perform the FD mode if there exists bidirectional traffic between it and its users. We assume that FD users/BSs use the same resource blocks to receive and transmit their data at the same time. Each resource block of a BS is only allocated to one of the users associating with the BS. Namely, if there are multiple users associating with the same BS, they cannot simultaneously share the same resource blocks.

Without loss of generality, consider a typical user $U_0$ located at the origin and our following location-dependent expressions and analyses will be based on the location of the typical user. Suppose all users adopt the following generalized user association (GUA) scheme to associate with their (downlink/uplink) BS\footnote{In order to simplify the notations in this paper, our following location-dependent expressions and analyses will be based on the location of typical user $U_0$ since the Slinvyak theorem shows that the statistical properties of a homogeneous PPP evaluated at any particular point are the same as those evaluated at other locations in the network\cite{DSWKJM13}. Also, we will study the scenario in which users can decouple their downlink and uplink BSs, i.e., the downlink BS and uplink BS could be different for the users. This decoupled user association can be achieved by adopting different user association functions for downlink and uplink.}:
\begin{align}\label{Eqn:GUA}
X_*\defn \arg\Psi_*(\|X_*\|)=\arg\sup_{m,i:X_{m,i}\in \mathcal{X}}\Psi_{m,i}(\|X_{m,i}\|),
\end{align}
where $X_*$ denotes the BS associated by typical user $U_0$, $\mathcal{X}\defn\bigcup_{m=1}^M\mathcal{X}_m$, $\|Y_i-Y_j\|$ denotes the Euclidean distance between nodes $Y_i$ and $Y_j$ for $i\neq j$, $\Psi_{m,i}:\mathbb{R}_{++}\rightarrow\mathbb{R}_+$ is called the user association function of BS $X_{m,i}$, $\Psi_*(\cdot)\in\{\Psi_{m,i}: m\in\mathcal{M}, i\in\mathbb{N}_+\}$ is the user association function of BS $X_*$, and $\Psi_*(\|X_*\|)\defn \sup_{m,i:X_{m,i}\in\mathcal{X}}\Psi_{m,i}(\|X_{m,i}\|)$. All $\Psi_{m,i}$'s are assumed to be a monotonic and bijective decreasing function. Furthermore, if they are random, they are i.i.d. for the same subscript $m$ and are independent for different subscripts $m$ and $i$. In the following analysis, we will use the following power-law-based function as the user association function $\Psi_{m,i}(\cdot)$ for BS $X_{m,i}$:
\begin{align}\label{Eqn:PowLawUserFun}
\Psi_{m,i}(x) = \frac{\psi_{m,i}}{x^{\alpha}},
\end{align}
where $\psi_{m,i}>0$ is the tier-$m$ random bias and $\alpha>2$ is called the path loss exponent. Although this GUA scheme is in principle the same as the user association scheme with constant biases in the literature, it is more general and can cover many existing based user association schemes. For example, if $\psi_{m,i}=1$ and we thus have $\Psi_{m,i}(\|X_{m,i}\|)=\|X_{m,i}\|^{-\alpha}$ that only characterizes the path loss between BS $X_{m,i}$ and typical user $U_0$, then users will associate with their nearest BS. In this case, the GUA scheme is essentially the nearest BS association (NBA) scheme \cite{CHLLCW1502}. If $\psi_{m,i}=P_m$ and $P_m$ is the transmit power of a tier-$m$ BS, then we have $\Psi_{m,i}(\|X_{m,i}\|)=P_m\|X_{m,i}\|^{-\alpha}$ that makes users associate with a BS that provides them with the maximum mean received power. In this case, the GUA scheme is called the mean maximum received-power association (MMPA) scheme \cite{CHLLCW16,CHLKLFONG16,CHLHCT1701}. Since different user association schemes induce different statistical properties of the signal-to-interference ratio (SIR) at receivers, in the following subsection we will first introduce the Laplace transform of an ``incomplete" Poisson shot-noise process that can be applied to model and analyze the interference in the sequel of the rate analysis. In addition, Table \ref{Tab:Notation} lists the notations of main variables, symbols and functions used in this paper.

\begin{table}
	\caption {NOTATION OF MAIN VARIABLES, SYMBOLS AND FUNCTIONS} \label{Tab:Notation}
		\begin{tabular}{ |c|c||c|c|}
			\hline
			Symbol & Meaning  & Symbol & Meaning \\ 
			\hline
			$U_j$ & User $j$ and its location & $\delta^c(n)$ & The complement of the Dirac delta function $\delta(n)$ \\
			$\mu$ &  User intensity & $F_{Z}(x) $ ($f_Z(x)$) & CDF (PDF) of Random Variable (RV) Z \\ 
			$Q$ & Transmit power of users & $\mathcal{L}_Z(\cdot)$ & Laplace  transform operator of RV $Z$ \\ $X_{m,i}$ & BS $i$ in the $m$th tier and its location & $\vartheta_m$ &  Tier-$m$ user association probability \\ $\lambda_m$ & Tier- $m$ BS intensity & $L_m$ & Tier-$m$ cell load \\  $\psi_{m,i}$ & Tier- $m$ BS $i$ association bias & $\rho_m$ & Tier-$m$ non-void probability \\$\|Y_i-Y_j\|$  & Distance between nodes $Y_i$ and $Y_j$ & $\epsilon_0 (\epsilon_*)$ & Self-interference suppression factor of user (BS) \\ $\alpha>2$ & Path loss exponent & $\dl$ ($\ul$)& Superscript for \textit{downlink} (\textit{uplink})\\ $X_*$ & The BS associated by typical user $U_0$ & $\gamma^{\dl}_{0}$ & The SIR of typical user $U_0$ \\ $P_m$ &  Transmit power of a tier-$m$ BS & $\gamma^{\ul}_{*}$& Full-duplex SIR of BS $X_*$\\ $H_{m,i}$ &  Channel gain from BS $X_{m,i}$ to user $U_0$ &
			$I^{\dl}_{\mathcal{X}}(I^{\ul}_{\mathcal{X}})$& Interference from all non-void BSs \\ $\breve{H}_{m,i}$ & Channel gain from BS $X_{m,i}$ to BS $X_*$ & $I^{\dl}_{\mathcal{U}}(I^{\ul}_{\mathcal{U}})$ & FD interference from all scheduled FD users  \\$G_{j}$ & Channel gain from user $U_j$ to user $U_0$ & $C^{\dl}_{\nu, FD}(C^{\ul}_{\nu, FD})$& Downlink (uplink) rate of an FD user \\$\breve{G}_j$ & Channel gain from user $U_j$ to BS $X_*$ & $a \gtrapprox b$ & $b$ is a tight lower bound on $a$ 
			\\$\mathsf{E}$ & Exponential RV with unit mean & $\Gamma(n,x)$&  Gamma RV with shape parameter $n$ and rate $x$ \\	$D_j\in\{0,1\}$ &$D_j=1$ if user $U_j$ is an FD user & $\nu$ & Probability of user $D_j$ being an FD user \\
			\hline
		\end{tabular}
\end{table}

\subsection{The Laplace Transform of Incomplete Poisson Shot-Noise Processes}
Consider a homogeneous PPP $\mathcal{Y}$ of intensity $\lambda_{\mathcal{Y}}$ and it can be written as $\mathcal{Y}\defn\{Y_i\in\mathbb{R}^2: i\in\mathbb{N}\}$. The $n$th-incomplete Poisson shot-noise process of $\mathcal{Y}$ is defined as
\begin{align}\label{Eqn:ImpPoissonNoise}
\mathfrak{I}_n\defn\sum_{i:Y_i\in\mathcal{Y}} W_{n+i} \xi\left(\|Y_{n+i}\|^2\right),\quad n\in\mathbb{N},
\end{align}
where $Y_{n+i}$ denotes the $(n+i)$-th nearest point in $\mathcal{Y}$ to the origin, $W_{n+i}$ is a random variable (RV) associated with $Y_i$, all $W_{n+i}$'s are i.i.d., $\xi:\mathbb{R}_{++}\rightarrow \mathbb{R}_+$ is a real-valued bijective function and its inverse is denoted by $\xi^{-1}(\cdot)$. The Laplace transform of a non-negative RV $Z$ is defined as
\begin{align*}
\mathcal{L}_Z(s) \defn\mathbb{E}\left[e^{-sZ}\right],\,\, s>0,
\end{align*}
and the Laplace transform of $\mathfrak{I}_n$ is shown in the following lemma.
\begin{lemma} \label{Lem:LapTransShotNoise}
If $\xi(\cdot)$ is a non-increasing and separable function\footnote{In this paper, a function $f(x_1,x_2,\ldots,x_n)$ is said to be separable if $f(x_1,x_2,\ldots,x_n)=\prod_{i=1}^{n}f(x_i)$ and its inverse function is also separable, i.e., $f^{-1}(x_1,\ldots,x_n)=\prod_{i=1}^{n} f^{-1}(x_i)$.} and $\mathbb{E}[\xi^{-1}(Z)]<\infty$ for any nonnegative RV $Z$, the Laplace transform of the $n$th-incomplete Poisson shot-noise process defined in \eqref{Eqn:ImpPoissonNoise} can be explicitly expressed as
\begin{align}
\mathcal{L}_{\mathfrak{I}_n}(s) =  \mathcal{L}_{\Xi_{\delta^c(n)}(\mathsf{E},\|Y_n\|^2,sW)}\left(\pi\lambda_{\mathcal{Y}}\right),\label{Eqn:LapTransShotNoise}
\end{align}
where  $\mathsf{E}\sim\exp(1)$ is an exponential RV with unit mean and variance, $\|Y_n\|^2\sim\Gamma(n,\pi\lambda_{\mathcal{Y}})$ is a Gamma RV with shape parameter $n\in\mathbb{N}_+$ and rate parameter $\pi\lambda_{\mathcal{Y}}$ (i.e., the pdf of $\|Y_n\|^2$ is $f_{\|Y_n\|^2}(y) = \frac{(\pi\lambda_{\mathcal{Y}})^ny^{n-1}e^{-\pi\lambda_{\mathcal{Y}}y}}{(n-1)!} $) , and function $\Xi _{\delta^c(n)}(x,y,z)$ is defined as
\begin{align}
\Xi_{\delta^c(n)}(x,y,z) \defn \mathbb{E}\left[\xi^{-1}\left(\frac{x}{z}\right)\right]+\delta^c(n)y\left[\int^{1}_{0}\mathcal{L}_z(\xi(yv)) \dif v-1\right]\label{Eqn:XiFun}
\end{align}
in which $\delta^c(n)\defn 1-\delta(n)$ is called the complement of the Dirac delta function $\delta(n)$.
\end{lemma}
\begin{IEEEproof}
See Appendix \ref{App:ProofLapTransShotNoise}.
\end{IEEEproof}

The result in Lemma \ref{Lem:LapTransShotNoise} is very general and it can be largely simplified in some special cases of $\xi(\cdot)$ and $n$. For instance, when $n=0$, we have a ``complete" Poisson shot-noise process and \eqref{Eqn:LapTransShotNoise} in this case reduces to 
\begin{align}\label{Eqn:ComShotNoisePro}
\mathcal{L}_{\mathfrak{I}_0}(s) =\Xi_0(\mathsf{E},0,sW)= \exp\left(-\pi\lambda_{\mathcal{Y}}\mathbb{E}\left[\xi^{-1}\left(\frac{\mathsf{E}}{sW}\right)\right]\right),
\end{align} 
and the pdf of $\mathfrak{I}_0$ can be obtained by finding the inverse Laplace transform of \eqref{Eqn:ComShotNoisePro}. A typical example that the pdf of $\mathfrak{I}_0$ can be found in closed-form is the case of  $\mathcal{L}_{\mathfrak{I}_0}(s)$ with $\xi(x) = x^{-2}$. For this case, we have $\mathcal{L}_{\mathfrak{I}_0}(s) = \exp\left(-\pi^{\frac{3}{2}}\lambda_{\mathcal{Y}}\mathbb{E}\left[\sqrt{W}\right] \sqrt{s}\right)$
and its inverse Laplace transform (i.e., the pdf of $\mathfrak{I}_0$) can be found as
\begin{align}
f_{\mathfrak{I}_0}(x) = \frac{\pi\lambda_{\mathcal{Y}}\mathbb{E}\left[\sqrt{W}\right]}{2\sqrt{ x^3}}\exp\left(-\frac{\pi^3 \lambda^2_{\mathcal{Y}}(\mathbb{E}[\sqrt{W}])^2}{4x}\right),
\end{align}
which is the same as the result shown in \cite{MH12}. For other cases of $n\geq 1$, the explicit expression of $\mathcal{L}_{\mathfrak{I}_n}$ in \eqref{Eqn:LapTransShotNoise} can be applied to evaluate the transmission performances of a user in different contexts such as user association, interference cancellation and BS coordination \cite{PXCHLJGA13,CHLHCT1701}, etc. We will need \eqref{Eqn:LapTransShotNoise} to facilitate the rate analyses in Section \ref{Sec:RateAnalysis}.

In the following subsection, some important statistical properties related to the GUA scheme with $\Psi_{m,i}(x)$ in \eqref{Eqn:PowLawUserFun} are introduced and they are the foundations of analyzing the SIR-related performance metrics in the HetNet, such as coverage and link rate.

\subsection{Statistical Properties for Generalized User Association (GUA)}\label{SubSec:UserAssociation}
In this subsection, some of the statistical properties related to the GUA scheme in \eqref{Eqn:GUA} are introduced. First, the distribution of the maximum user association function in \eqref{Eqn:GUA} with the user association function in \eqref{Eqn:PowLawUserFun}, i.e., the cumulative density function (CDF) of $\Psi_*(\|X_*\|)$ in \eqref{Eqn:GUA} with $\Psi_{m,i}(x)=\psi_{m,i} x^{-\alpha}$ in \eqref{Eqn:PowLawUserFun}, can be found by using Theorem 1 in our previous work \cite{CHLKLFONG16}. Its explicit result and tier-$m$ association probability that is obtained by  Theorem 2 in \cite{CHLKLFONG16} are summarized in the following lemma.
\begin{lemma}\label{Lem:StatUserAssociation}
Suppose all users adopt the GUA scheme in \eqref{Eqn:GUA} with the user association function in \eqref{Eqn:PowLawUserFun} to associate their (downlink or uplink) BS. The CDF of $\Psi_*(\|X_*\|) $ can be shown as
	\begin{align}\label{Eqn:CDFPowerLawMaxUserAss}
	F_{\Psi_*(\|X_*\|)}(x) = \exp\left(-\pi x^{-\frac{2}{\alpha}}\sum_{m=1}^{M} \lambda_m  \mathbb{E}\left[\psi^{\frac{2}{\alpha}}_{m}\right]\right).
	\end{align}
\end{lemma}
\noindent The CDF in \eqref{Eqn:CDFPowerLawMaxUserAss} essentially indicates that it can be equivalently found by assuming there is a PPP of intensity $\sum_{m=1}^{M}\lambda_m\mathbb{E}\left[\psi^{\frac{2}{\alpha}}_m\right]$ and all BSs in this PPP use the same user association function $\Psi_{m,i}(x)=x^{-\alpha}$, which is an unbiased power-law function of $x$. In other words, the distance between the origin and the nearest point in this PPP has the same distribution as $(\Psi_*(\|X_*\|))^{-\frac{1}{\alpha}}$ because $\mathbb{P}\left[(\Psi_*(\|X_*\|))^{-\frac{1}{\alpha}}\geq x\right]=\exp\left(-\pi x^2\sum_{m=1}^{M}\lambda_m\mathbb{E}\left[\psi^{\frac{2}{\alpha}}_m\right]\right)$ is the complement CDF (CCDF) of the distance from the origin to the nearest point in the PPP of intensity $\sum_{m=1}^{M}\lambda_m\mathbb{E}\left[\psi^{\frac{2}{\alpha}}_m\right]$. Also, the tier-$m$ cell load based on Lemma 1 in \cite{CHLKLFONG16}, denoted by $L_m$, can be written as
\begin{align}\label{Eqn:Tier-mCellLoad}
L_m = \frac{\mu}{\lambda_m}\vartheta_m,
\end{align}  
where $\vartheta_m$	is called tier-$m$ association probability given by
\begin{align}\label{Eqn:Tire-mUserAssProb}
\vartheta_m =\frac{\lambda_m\mathbb{E}\left[\psi^{\frac{2}{\alpha}}_m\right]}{\sum_{k=1}^{M}\lambda_k\mathbb{E}\left[\psi^{\frac{2}{\alpha}}_k\right]},
\end{align}
which is the probability that a user associates with a tier-$m$ BS. Furthermore, if the GUA scheme is adopted, by using Lemma 1 in \cite{CHLKLFONG16} the tier-$m$ non-void probability that a tier-$m$ BS is associated by at least one user, denoted by $\rho_m$, can be found  as  
\begin{align}\label{Eqn:NonvoidProbTier-m}
\rho_m=1-\underbrace{\left(1+\frac{\mu\vartheta_m}{\zeta_m\lambda_m}\right)^{-\zeta_m}}_{\text{Void Probability of a tier-$m$ BS }}=1-\left(1+\frac{L_m}{\zeta_m}\right)^{-\zeta_m},
\end{align}
where $\zeta_m\defn\frac{7}{2}\mathbb{E}\left[\psi^{\frac{2}{\alpha}}_m\right]\mathbb{E}\left[\psi^{-\frac{2}{\alpha}}_m\right]$, i.e., $1-\rho_m$ is the tier-$m$ void probability that a tier-$m$ BS is not associated by any users. Obviously, $\rho_m$ is small (or $1-\rho_m$ is not small) whenever the user intensity is not large relative to the total intensity of all the BSs. A smaller $\rho_m$ indicates that the HetNet has lesser interference since the void BSs do not generate any interference and many prior works on the interference modeling in a HetNet overlook this important issue. Later, we will see that the results in \eqref{Eqn:CDFPowerLawMaxUserAss}-\eqref{Eqn:NonvoidProbTier-m} can be applied to explicitly characterize the FD link rates which are defined based on the signal-to-interference ratio (SIR) model introduced in the following subsection.

\subsection{Full-Duplex SIR Model for Decoupled GUA}

 \begin{figure} 	
 	\centering\includegraphics[width=6.5 in,height=2.25in]{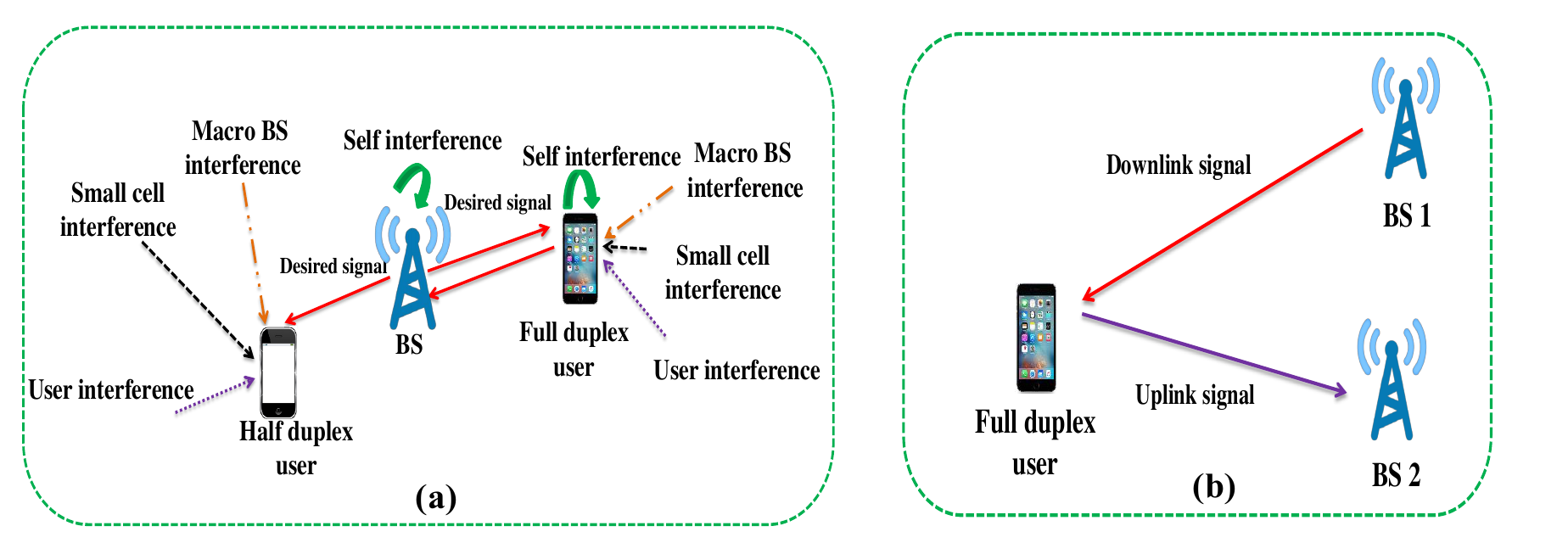}
	\caption{(a) An illustration of the HD and FD transmission scenarios for users and their tagged BS. The BS and its serving user perform the FD mode while they have bidirectional traffic and otherwise they perform the HD mode. Note that the HD and FD users suffer different interferences. (b) An illustration of the decoupled user association scenario in which the FD user associates with the different downlink and uplink BSs.}
	\label{Fig:SystemModel}
\end{figure}

For the full-duplex HetNet considered in this paper, we will study the decoupled user association scenario in which users adopt different uplink and downlink user association functions in \eqref{Eqn:GUA} for associating with a uplink BS as well as a downlink BS, that is the decoupled GUA (DGUA) scheme shown in the following\footnote{Throughout this paper, the variables/symbols with superscript ``$\dl$" indicate that they are in the \textit{downlink} context, whereas the variables/symbols with superscript ``$\ul$" mean that they are in the \textit{uplink} context. For example, here $\psi^{\dl}_{m,i}$ denotes the user association bias of downlink BS $X^{\dl}_{m,i}$ whereas  $\psi^{\ul}_{m,i}$ represents the user association bias of uplink BS $X^{\ul}_{m,i}$.}:
\begin{align}\label{Eqn:DwUpUserAss}
X_*=\begin{cases}
X^{\dl}_* = \arg\sup_{m,i:X_{m,i}\in\mathcal{X}} \frac{\psi^{\dl}_{m,i}}{\|X_{m,i}\|^{\alpha}}, &\text{ for downlink} \\
X^{\ul}_*=\arg\sup_{m,i:X_{m,i}\in\mathcal{X}} \frac{\psi^{\ul}_{m,i}}{\|X_{m,i}\|^{\alpha}}, &\text{ for uplink}
\end{cases},
\end{align}
where $X^{\dl}_*$ ($X^{\ul}_*$) is the downlink (uplink) BS associated by typical user $U_0$ and $\psi^{\dl}_{m,i}(\cdot)$ ($\psi^{\ul}_{m,i}(\cdot)$) is the downlink (uplink) user association bias. An illustration of the downlink-uplink decoupled transmission scenario is shown in Fig. \ref{Fig:SystemModel} for an FD user and an HD user in a two-tier HetNet where case (a) indicates the scenario of coupled user association and case (b) shows the scenario of decoupled user association. Due to full-duplex, the FD user would receive different interferences from other BSs and FD users in the HetNet in addition to its own self-interference. 

Since an FD user suffers from different interferences in the network as shown in Fig. \ref{Fig:SystemModel}, we need to specify an appropriate full-duplex SIR model for the following rate analysis. Let $D_j\in\{0,1\}$ for all $j\in\mathbb{N}$ be a Bernoulli RV that is unity if there exists FD traffic between user $j$ and its tagged BS and zero if there is only downlink traffic from a BS to its user $U_j$. Namely, $D_j$ indicates whether user $U_j$ is an FD user. Assume the uplink traffic patterns of all users are independent and all BSs independently make their traffic scheduling decision so that the scheduled FD users form a thinning PPP that is a subset of set $\mathcal{U}$. The SIR of typical user $U_0$ can be written as
\begin{align}
\gamma^{\dl}_{0}=\frac{H_*P_*\|X^{\dl}_*\|^{-\alpha}}{I_{0}+\epsilon_0 Q D_0}, \label{Eqn:DownlinkUserSIR}
\end{align}
where $H_*\in\mathbb{R}_{++}$ is the fading channel gain of BS $X^{\dl}_*$, $P_*\in\{P_1,P_2,\ldots,P_M\}$ is the transmit power of BS $X^{\dl}_*$, $P_m$ is the transmit power of the tier-$m$ BSs, $Q$ is the transmit power of users,  $\epsilon_0\in[0,1]$ denotes the self-interference suppression factor of users ($\epsilon_0=0$ for canceling the self-interference completely; otherwise  $\epsilon_0\neq0$.), $\epsilon_0 Q$ denotes the residual fraction of self-interference $Q$, and $I_{0}$ is the interference given by
\begin{align}\label{Eqn:DownLinkInterference}
I_{0}\defn I^{\dl}_{\mathcal{X}}+I^{\dl}_{\mathcal{U}} = \underbrace{\sum_{m,i:X_{m,i}\in\mathcal{X}\setminus X_*}V^{\dl}_{m,i}\frac{P_mH_{m,i}}{\|X_{m,i}\|^{\alpha}}}_{I^{\dl}_{\mathcal{X}}}+\underbrace{\sum_{j:U_j\in\mathcal{U}}\frac{QD_jG_j}{\|U_j\|^{\alpha}}}_{I^{\dl}_{\mathcal{U}}},
\end{align}
where $ I^{\dl}_{\mathcal{X}}$ denotes the interference from all non-void BSs, $I^{\dl}_{\mathcal{U}}$ is the FD interference from all scheduled FD users in set $\mathcal{U}$, $V^{\dl}_{m,i}\in\{0,1\}$ is a Bernoulli RV that is zero when BS $X_{m,i}$ is void in the downlink and one otherwise, $H_{m,i}$ denotes the (fading and/or shadowing) channel gain from BS $X_{m,i}$ to typical user $U_0$ (All $H_{m,i}$'s are independent for all $m\in\mathcal{M}$ and $i\in\mathbb{N}_+$ and they are i.i.d. for the same subscript $m$.), and $G_j$ denotes the fading channel gain from $U_j$ to $U_0$ (All $G_j$'s are i.i.d. for all $j\in\mathbb{N}_+$). Note that all $V^{\dl}_{m,i}$'s may not be completely independent due to downlink user association, but the correlations between them are fairly weak in general \cite{CHLLCW1502}.

When BS $X_*$ needs to serve FD users and operate in the FD mode, we assume that its downlink and uplink channels are reciprocal. Thus, the full-duplex SIR of BS $X_*$ is written as
\begin{align}\label{Eqn:UplinkSIR}
\gamma^{\ul}_{*} = \frac{QH_*\|X^{\ul}_*\|^{-\alpha}}{(I_*+\epsilon_* P_*)},
\end{align}
where $\epsilon_*\in[0,1]$ is the self-interference suppression factor of BS $X^{\ul}_*$, $\epsilon_* P_*$ denotes the residual self-interference of BS $X^{\ul}_*$, $I_*$ denotes the interference received by BS $X^{\ul}_*$ and it is given by
\begin{align}\label{Eqn:UplinkInterference}
I_*\defn I^{\ul}_{\mathcal{X}}+I^{\ul}_{\mathcal{U}} = \underbrace{\sum_{m,i:X_{m,i}\in \mathcal{X}\setminus X^{\ul}_*}V^{\ul}_{m,i}\frac{P_m\breve{H}_{m,i}}{\|X^{\ul}_*-X_{m,i}\|^{\alpha}}}_{I^{\ul}_{\mathcal{X}}}+\underbrace{\sum_{j:U_j\in\mathcal{U}}\frac{QD_j\breve{G}_j}{\|X^{\ul}_*-U_j\|^{\alpha}}}_{I^{\ul}_{\mathcal{U}}},
\end{align}  
where $I^{\ul}_{\mathcal{X}}$ denotes the interference from all non-void BSs, $I^{\ul}_{\mathcal{U}}$ is the FD interference from all scheduled FD users, $V^{\ul}_{m,i}\in\{0,1\}$ is a Bernoulli RV that is zero if BS $X_{m,i}$ is void and one otherwise, $\breve{H}_{m,i}$ that has the same distribution as $H_{m,i}$ for all $m\in\mathcal{M}$ and $i\in\mathbb{N}_+$ denotes the channel gain from $X_{m,i}$ to $X_*$ and $\breve{G}_j$ that has the same distribution as $G_j$ for all $j\in\mathbb{N}_+$
is the channel gain from $U_i$ to $X_*$. The full-duplex SIR models above for downlink and uplink can be used to characterize the rate-optimal user association scheme that is introduced in the following subsection. Also, all $V^{\dl}_{m,i}$'s may not be completely independent owing to uplink user association, but their correlations are in general very weak \cite{CHLLCW1502}.

\subsection{Decoupled Rate-Optimal User Association}

Consider the DGUA scheme in \eqref{Eqn:DwUpUserAss} and we designate its $\Psi^{\dl}_{m,i}(\cdot)$  and $\Psi^{\ul}_{m,i}(\cdot)$ as 
\begin{align}\label{Eqn:RateOptUserAssFun}
\Psi_{m,i}(\|X_{m,i}\|) =\begin{cases}
\Psi^{\dl}_{m,i}(\|X_{m,i}\|)=\log\left[1+\gamma^{\dl}_{m,i}(\|X_{m,i}\|)\right], &\text{ for downlink}\\
\Psi^{\ul}_{m,i}(\|X_{m,i}\|)=\log\left[1+\gamma^{\dl}_{m,i}(\|X_{m,i}\|)\right],&\text{ for uplink}
\end{cases},
\end{align}
where $\gamma^{\dl}_{m,i}(\|X_{m,i}\|)$ is the downlink SIR (BS $X_{m,i}$ is the transmitter and typical user $U_0$ is the receiver) and $\gamma^{\ul}_{m,i}(\|X_{m,i}\|)$ is the uplink SIR (BS $X_{m,i}$ is the receiver and typical user $U_0$ is the transmitter). The DGUA scheme with the decoupled user association function in \eqref{Eqn:RateOptUserAssFun} is called the decoupled ``rate-optimal" association (DROA) scheme since it selects the BS that can make users achieve the maximum downlink and uplink rates among all BSs. The DROA scheme can be simplified as shown in the following lemma.
\begin{lemma}\label{Lem:DROA}
If the DGUA scheme in \eqref{Eqn:DwUpUserAss} uses the following decoupled user association function
\begin{align}\label{Eqn:UserAssFunROA}
\Psi_{m,i}(\|X_{m,i}\|)=\begin{cases}
\Psi^{\dl}_{m,i} = \frac{P_mH_{m,i}}{\|X_{m,i}\|^{\alpha}},&\text{ for downlink}\\
\Psi^{\ul}_{m,i} = \frac{\breve{H}_{m,i}}{\|X_{m,i}\|^{\alpha}},&\text{ for uplink}
\end{cases},
\end{align}
then it is the same as the DROA scheme defined in \eqref{Eqn:RateOptUserAssFun}. 
\end{lemma}
\begin{IEEEproof}
See Appendix \ref{App:ProofDROA}.
\end{IEEEproof}

Lemma \ref{Lem:DROA} gives us an important insight into how an FD user should associate with its downlink and uplink BSs in order to maximize its FD link rate; that is, for the downlink an FD user should select the BS that provides the maximum received signal power to it, whereas for the uplink the FD user should associate with a BS that has the maximum channel gain from the FD user to it. Although the DROA scheme can achieve the maximum downlink and uplink rates, respectively, it may not be easily implemented in practice since it needs to instantaneously catch up the fading variations of all channels between a user and all BSs. In practice, users are more likely to get the means of the fading channel gains so that $\Psi_{m,i}(\|X_{m,i}\|)$ in \eqref{Eqn:UserAssFunROA} is modified as
\begin{align}\label{Eqn:UserAssFunROAMod}
\Psi_{m,i}(\|X_{m,i}\|)=\begin{cases}
\Psi^{\dl}_{m,i} = \frac{P_m\mathbb{E}[H_{m}]}{\|X_{m,i}\|^{\alpha}},&\text{for downlink}\\
\Psi^{\ul}_{m,i} = \frac{\mathbb{E}[\breve{H}_m]}{\|X_{m,i}\|^{\alpha}},&\text{for uplink}
\end{cases},
\end{align}
which is thus called the modified DROA (MDROA) scheme. Obviously, the downlink and uplink rates achieved by MDROA are inferior to those achieved by DROA since MDROA does not exploit the channel fading diversity among all BSs. In sum, Lemma \ref{Lem:DROA} reveals two important facts: (i) In general, an FD user in a multi-tier HetNet should associate with different downlink and uplink BSs in order to improve their bidirectional rates. (ii) Without using the ROA scheme in \eqref{Eqn:UserAssFunROA}, an FD user only can achieve suboptimal bidirectional rates. In other words, using other user association schemes, e.g., the NBA scheme which is the most popular scheme used in the literature, cannot achieve the rate optimality of a HetNet with different transmit powers and channel fading statistics in different tiers. In the following section, we will study how much link rate can be achieved by the DGUA, DROA and other schemes. 

\section{Rate Analysis for the DGUA Scheme}\label{Sec:RateAnalysis}
In this section, we will first study the link rate achieved by the DGUA scheme in \eqref{Eqn:DwUpUserAss}. Our primary goal here is to generally characterize the downlink and uplink rates of a user so that we can know how different downlink and uplink user association schemes, channel models and imperfect self-interference cancellation influence the rate performance in an FD HetNet, which gives us some insight into how to boost the overall network throughput. Next, we will derive and analyze the downlink and uplink link rates while the DROA scheme is adopted. These derived rate results not only shed light on how much link rate can be fundamentally attained by an FD user, but also indicate how users and BSs should schedule their uplink and downlink traffic in order to maximize their bidirectional rates. 

\subsection{Analysis of the Downlink and Uplink rates with DGUA}\label{SubSec:rateAnalysis}
In this subsection, we study the maximum (achievable) downlink and uplink rates of an FD user while considering the entire spectrum resource of a BS is given to one user at a time and the DGUA scheme in \eqref{Eqn:DwUpUserAss} is adopted. Recall that $D_j\in\{0,1\}$ is FD traffic pattern parameter between $U_j$ and its tagged BS, i.e. user $U_j$ is an FD user if $D_j=1$ and it is an HD user if $D_j=0$.  Let us define $\nu\defn \mathbb{P}[D_j=1]$ for all $j\in\mathbb{N}$ be the FD traffic pattern parameter between a user and its tagged BS so that $\nu$ has the physical meaning of how likely FD traffic happens between a user and its tagged BS. The downlink rate of an FD user is defined as
\begin{align}\label{Eqn:DefnDownlinkRate}
C^{\dl}_{\nu, FD} \defn \mathbb{E}\left[\log \left(1+\gamma^{\dl}_0\right)\right],\quad\text{(nats/Hz)},
\end{align}
whereas the uplink rate of an FD user is defined as
\begin{align}\label{Eqn:DefnUplinkRate}
C^{\ul}_{\nu, FD}\defn \mathbb{E}\left[\log(1+\gamma_*^{\ul})\right], \quad\text{(nats/Hz)}.
\end{align}
Note that $C^{\dl}_{\nu,FD}$ and $C^{\ul}_{\nu,HD}$ both are not independent because the uplink BS $X^{\ul}_*$ and  the downlink BS $X^{\dl}_*$ are found in the same BS set so that the downlink distance $\|X^{\dl}_*\|$ in  $C^{\dl}_{\nu,FD}$ and the uplink distance $\|X^{\ul}_*\|$ in $C^{\ul}_{\nu,HD}$ are not independent. In addition, $C^{\dl}_{\nu,FD}$ and $C^{\ul}_{\nu,HD}$ both depend on parameter $\nu$, as shown in the following position.  
\begin{proposition}\label{Prop:FDLinkrate}
If all FD users adopt the DROA scheme in \eqref{Eqn:RateOptUserAssFun} to associate with their downlink and uplink BSs, the downlink rate in \eqref{Eqn:DefnDownlinkRate} is tightly lower-bounded by  
\begin{align}
C^{\dl}_{\nu, FD}\gtrapprox \sum_{m=1}^{M}\vartheta^{\dl}_m\int_{0^+}^{\infty}\int_{0}^{\infty}\frac{\left[1-\mathcal{L}_{H_m/\psi^{\dl}_m}(s)\right]\dif y\dif s}{s\exp\left\{\frac{s\epsilon_0 Q}{\pi P_m \widetilde{\lambda}^{\dl}}y^{\frac{\alpha}{2}}+y [\widetilde{\Xi}^{\dl}_m(s)+1]\right\}},\label{Eqn:DLrate}
\end{align}
where $a \gtrapprox b$ means that $b$ is a tight lower bound on $a$, $\vartheta^{\dl}_m\defn\frac{\lambda_m\mathbb{E}\left[(\psi^{\dl}_m)^{\frac{2}{\alpha}}\right]}{\sum_{k=1}^{M}\lambda_k\mathbb{E}\left[(\psi^{\dl}_k)^{\frac{2}{\alpha}}\right]}$ as defined in \eqref{Eqn:Tire-mUserAssProb} is the probability that a user associates with a tier-$m$ BS in the downlink, $\widetilde{\Xi}^{\dl}_m(s)$ is defined as
\begin{align}\label{Eqn:DLXiFun}
\widetilde{\Xi}^{\dl}_m(s) \defn \sum_{k=1}^{M}\vartheta^{\dl}_k\rho^{\dl}_k\Xi_1\left(\mathsf{E},1,\frac{sP_kH_k}{P_m\psi^{\dl}_k}\right)+\nu\Gamma\left(1-\frac{2}{\alpha}\right)\left(\frac{\sum_{k=1}^{M}\lambda_k\rho^{\ul}_k}{\widetilde{\lambda}^{\dl}}\right)\mathbb{E}\left[\left(\frac{sQG}{P_m}\right)^{\frac{2}{\alpha}}\right],
\end{align}
in which $\Xi_1(\cdot)$ is defined in \eqref{Eqn:XiFun},  Gamma function $\Gamma(x)\defn\int_{0}^{\infty} t^{x-1} e^{-t} \dif t$, $\widetilde{\lambda}^{\dl}\defn\sum_{k=1}^{M}\lambda_k\mathbb{E}[(\psi^{\dl}_k)^{\frac{2}{\alpha}}]$ and $\mathsf{E}\sim\exp(1)$.

For the uplink rate in \eqref{Eqn:DefnUplinkRate}, its tight lower bound can be characterized as
\begin{align}\label{Eqn:ULrate}
C^{\ul}_{\nu, FD}\gtrapprox  \sum_{m=1}^{M}\vartheta^{\ul}_m\int_{0^+}^{\infty}\int_{0}^{\infty}\frac{\left[1-\mathcal{L}_{H_m/\psi^{\ul}_m}(s)\right]\dif y\dif s}{s\exp\left\{\frac{s\epsilon_* P_m}{\pi Q \widetilde{\lambda}^{\ul}}y^{\frac{\alpha}{2}}+y [\widetilde{\Xi}^{\ul}(s)+1]\right\}},
\end{align}
where $\widetilde{\Xi}^{\ul}(s)$ is defined as
\begin{align}\label{Eqn:ULXiFun}
\widetilde{\Xi}^{\ul}(s) \defn \Gamma\left(1-\frac{2}{\alpha}\right)s^{\frac{2}{\alpha}}\left\{\sum_{k=1}^{M}\vartheta_k^{\ul}\rho^{\ul}_k\mathbb{E}\left[\left(\frac{P_kH_k}{Q\psi^{\ul}_k}\right)^{\frac{2}{\alpha}}\right]+\nu\left(\frac{\sum_{k=1}^{M}\lambda_k\rho^{\ul}_k}{\widetilde{\lambda}^{\ul}}\right)\mathbb{E}\left[G^{\frac{2}{\alpha}}\right]\right\}
\end{align}
with $\vartheta^{\ul}_m\defn\frac{\lambda_m\mathbb{E}\left[(\psi^{\ul}_m)^{\frac{2}{\alpha}}\right]}{\sum_{k=1}^{M}\lambda_k\mathbb{E}\left[(\psi^{\ul}_k)^{\frac{2}{\alpha}}\right]}$ and $\widetilde{\lambda}^{\ul}\defn\sum_{m=1}^{M}\lambda_k\mathbb{E}\left[(\psi^{\ul}_k)^{\frac{2}{\alpha}}\right]$.
\end{proposition}
\begin{IEEEproof}
See Appendix \ref{App:ProofFDLinkrate}.
\end{IEEEproof}

Even though the tight bounds on the downlink and uplink rates shown in Proposition \ref{Prop:FDLinkrate} are somewhat complex, they are very general and suited for any fading channel models, user association schemes and imperfect self-interference cancellation. Most importantly, they characterize the downlink and uplink void BSs that do not generate interferences so that they reveal how different cell loads induced by different user association schemes affect the link rate. To the best of our knowledge, \eqref{Eqn:DLrate} and \eqref{Eqn:ULrate} are the most general and accurate expressions with moderate complexity and their lower bounds are exactly achieved as the user intensity goes to infinity, i.e., considering the ``full load" case, $C^{\dl}_{\nu, FD}$ and $C^{\ul}_{\nu, FD}$ reduce to their lower bounds in \eqref{Eqn:DLrate} and \eqref{Eqn:ULrate} with $\widetilde{\Xi}^{\dl}_m(s)$ and $\widetilde{\Xi}^{\ul}(s)$ given by
\begin{align}
\widetilde{\Xi}^{\dl}_m(s)&= \sum_{k=1}^{M}\vartheta^{\dl}_k\Xi_1\left(\mathsf{E},1,\frac{sP_kH_k}{P_m\psi^{\dl}_k}\right)+\frac{\nu\lambda}{\widetilde{\lambda}^{\dl}}\Gamma\left(1-\frac{2}{\alpha}\right)\mathbb{E}\left[\left(\frac{sQG}{P_m}\right)^{\frac{2}{\alpha}}\right],\\
\widetilde{\Xi}^{\ul}(s)&=\Gamma\left(1-\frac{2}{\alpha}\right)s^{\frac{2}{\alpha}}\left\{\sum_{k=1}^{M}\vartheta^{\ul}_k\mathbb{E}\left[\left(\frac{P_kH_k}{Q\psi^{\ul}_k}\right)^{\frac{2}{\alpha}}\right]+\frac{\nu\lambda}{\widetilde{\lambda}^{\ul}}\mathbb{E}\left[G^{\frac{2}{\alpha}}\right]\right\},
\end{align}
where $\lambda\defn\sum_{k=1}^{M}\lambda_k$. 

For some special cases, \eqref{Eqn:DLrate} and \eqref{Eqn:ULrate} can be largely simplified, as discussed in the following:
\subsubsection{No Self-Interference} 
Suppose an FD user and its serving BS both can completely cancel their self-interferences. Thus, in this situation we have $\epsilon_0=\epsilon_*=0$ so that $C^{\dl}_{\nu, FD}$ and $C^{\ul}_{\nu, FD}$ reduce to
\begin{align}\label{Eqn:LinkrateNoSelfInt}
\hspace*{-0.15in}C^{\dl}_{\nu, FD}\gtrapprox \sum_{m=1}^{M}\vartheta^{\dl}_m\int_{0^+}^{\infty}\frac{\left[1-\mathcal{L}_{H_m/\psi^{\dl}_m}(s)\right]}{s\left[\widetilde{\Xi}^{\dl}_m(s)+1\right]}\dif s\,\,\text{  and  }\,\,C^{\ul}_{\nu, FD}\gtrapprox \sum_{m=1}^{M}\vartheta^{\ul}_m\int_{0^+}^{\infty}\frac{\left[1-\mathcal{L}_{H_m/\psi^{\ul}_m}(s)\right]}{s\left[\widetilde{\Xi}^{\ul}(s)+1\right]}\dif s,
\end{align}
and they have a much simpler form with a single integral. 

\subsubsection{Using DROA and No Self-Interference} In this case, using the DROA scheme in \eqref{Eqn:UserAssFunROA} makes \eqref{Eqn:LinkrateNoSelfInt} further reduce to
\begin{align}
C^{\dl}_{\nu, FD}\gtrapprox \int_{0^+}^{\infty}\frac{\left(1-e^{-s}\right)}{s\left[\widetilde{\Xi}^{\dl}_m(sP_m)+1\right]}\dif s\,\,\text{  and  }\,\,C^{\ul}_{\nu, FD}\gtrapprox \int_{0^+}^{\infty}\frac{\left(1-e^{-s}\right)}{s\left[\widetilde{\Xi}^{\ul}(s)+1\right]}\dif s\label{Eqn:LinkrateDROANoSelfInt}
\end{align}
in which $\widetilde{\Xi}^{\dl}_m(sP_m)$ and $\widetilde{\Xi}^{\ul}(s)$ reduce to
\begin{align}
\widetilde{\Xi}^{\dl}_m(sP_m) &= \Xi_1\left(\mathsf{E},1,s\right)\sum_{k=1}^{M}\vartheta^{\dl}_k\rho^{\dl}_k+\nu\Gamma\left(1-\frac{2}{\alpha}\right)\left(\frac{\sum_{k=1}^{M}\lambda_k\rho^{\ul}_k}{\widetilde{\lambda}^{\dl}}\right)\mathbb{E}\left[\left(sQG\right)^{\frac{2}{\alpha}}\right],\label{Eqn:DLXiFunDROANoSelf}\\
\widetilde{\Xi}^{\ul}(s) &= \Gamma\left(1-\frac{2}{\alpha}\right)s^{\frac{2}{\alpha}}\left\{\sum_{k=1}^{M}\vartheta^{\ul}_k\left(\frac{P_k}{Q}\right)^{\frac{2}{\alpha}}\rho^{\ul}_k+\nu\left(\frac{\sum_{k=1}^{M}\lambda_k\rho^{\ul}_k}{\widetilde{\lambda}^{\ul}}\right)\mathbb{E}\left[G^{\frac{2}{\alpha}}\right]\right\}, \label{Eqn:ULXiFunDROANoSelf}
\end{align}
respectively. Note that the results in \eqref{Eqn:LinkrateDROANoSelfInt} represent the maximum achievable downlink and uplink rates because Lemma \ref{Lem:DROA} has shown that DROA is able to achieve the maximum downlink and uplink rates at the same time. In other words, \textit{any decoupled and coupled user association schemes cannot outperform the DROA scheme in terms of the sum of the downlink and uplink rates}. However, the uplink and downlink maximum rates may not be easily achievable since the user association process in general may not be done within the channel coherence time, as pointed out before. Instead of using DROA, we can adopt the MDROA scheme to achieve the link rates that would be just slightly smaller than those achieved by DROA.  

\subsubsection{No Fading, Using MDROA and No Self-Interference} Since there is no fading in all channel, all channel gains are equal to unity and we thus have the link rates as show in \eqref{Eqn:LinkrateDROANoSelfInt} with $\widetilde{\Xi}^{\dl}_m(s)$ and $\widetilde{\Xi}^{\ul}(s)$ given by
\begin{align}
\widetilde{\Xi}^{\dl}_m(sP_m) &= \Xi_1\left(\mathsf{E},1,s\right)\sum_{k=1}^{M}\vartheta^{\dl}_k\rho^{\dl}_k+\nu\left(sQ\right)^{\frac{2}{\alpha}}\Gamma\left(1-\frac{2}{\alpha}\right)\left(\frac{\sum_{k=1}^{M}\lambda_k\rho^{\ul}_k}{\widetilde{\lambda}^{\dl}}\right),\label{Eqn:DLXiFunDROANoSelfNoFading}
\end{align}
and
\begin{align}
\widetilde{\Xi}^{\ul}(s) &= \Gamma\left(1-\frac{2}{\alpha}\right)\frac{s^{\frac{2}{\alpha}}}{\widetilde{\lambda}^{\ul}}\left\{\sum_{k=1}^{M}\lambda_k\rho^{\ul}_k\left[\left(\frac{P_k}{Q}\right)^{\frac{2}{\alpha}}+\nu\right]\right\}. \label{Eqn:ULXiFunDROANoSelfNoFading}
\end{align}
No that in the literature the ergodic link rate without fading cannot be tractably found in a neat form but it can be explicitly found by using the results in Proposition \ref{Prop:FDLinkrate}.

\subsubsection{Rayleigh Fading Channels, Using MDROA and No Self-Interference} In this case, all channel gains are i.i.d. exponential RVs with unit mean and variance. As such, we still have \eqref{Eqn:LinkrateDROANoSelfInt}, but its $\widetilde{\Xi}^{\dl}_m(s)$ and $\widetilde{\Xi}^{\ul}(s)$ reduce to
\begin{align}
\widetilde{\Xi}^{\dl}_m(sP_m) =& \left[\frac{s^{\frac{2}{\alpha}}}{\mathrm{sinc}(2/\alpha)}+\Gamma\left(1+\frac{2}{\alpha}\right)\left(\int_{0}^{1}e^{-\frac{s}{P_m}v^{-\frac{\alpha}{2}}}\dif v-1\right)\right]\left(\sum_{k=1}^{M}\vartheta^{\dl}_k\rho^{\dl}_k\right)\nonumber\\
&+\frac{\nu\left(sQ\right)^{\frac{2}{\alpha}}}{\mathrm{sinc}(2/\alpha)}\left(\frac{\sum_{k=1}^{M}\lambda_k\rho^{\ul}_k}{\widetilde{\lambda}^{\dl}}\right)
\end{align}
and
\begin{align}
\widetilde{\Xi}^{\ul}(s) =&\frac{s^{\frac{2}{\alpha}}}{\widetilde{\lambda}^{\ul}\mathrm{sinc}(2/\alpha)} \left\{\sum_{k=1}^{M}\lambda_k\rho^{\ul}_k\left[\left(\frac{P_k}{Q}\right)^{\frac{2}{\alpha}}+\nu\right]\right\},
\end{align}
where $\mathrm{sinc}(x)\defn\frac{\sin(\pi x)}{\pi x}$. Thus, we get very neat and tight lower bounds on $C^{\dl}_{\nu, FD}$ and $C^{\ul}_{\nu, FD}$. 

In addition to the great feature of generality in the tight bounds in \eqref{Eqn:DLrate} and \eqref{Eqn:ULrate}, there are two important implications that can be learned by inspecting $\widetilde{\Xi}^{\dl}_m(s)$ in \eqref{Eqn:DLXiFun} and $\widetilde{\Xi}^{\ul}(s)$ in \eqref{Eqn:ULXiFun}. First, we learn that\textit{ using large user association biases helps to suppress the self-interference and the FD interferences} because making $\widetilde{\lambda}^{\dl}=\sum_{k=1}^{M}\lambda_k\mathbb{E}\left[\left(\psi^{\dl}_k\right)^{\frac{2}{\alpha}}\right]$ and $\widetilde{\lambda}^{\ul}=\sum_{k=1}^{M}\lambda_k\mathbb{E}\left[\left(\psi^{\ul}_k\right)^{\frac{2}{\alpha}}\right]$ larger by increasing $\psi^{\dl}_k$'s and $\psi^{\ul}_k$'s reduces the denominators of $C^{\dl}_{\nu, FD}$ and $C^{\ul}_{\nu, FD}$. Second, \textit{offloading more traffic (or using larger user association biases) to the tiers with a higher intensity helps efficiently suppress the self-interference and the FD interferences, whereas deploying more BSs without using appropriate user association biases may not improve the link rates}.  In the following subsection, some numerical results are provided to validate the derived tight lower bounds on $C^{\dl}_{\nu, FD}$ and $C^{\ul}_{\nu, FD}$ and how they are affected by BS intensities, FD/HD transmission and imperfect self-interference cancellation.

\subsection{Numerical Results and Discussions}\label{SubSec:Simulationsrate}
\begin{table}[!t] 
	\caption {Network parameters for simulation}\label{tab:parameter} 
	\begin{center}
		\begin{tabular}{|c|c|c|} 
			\hline
			Parameter $\setminus$ BS Type (Tier \#) & Macrocell BS (1) & Small cell BS (2) \\  	
			\hline Transmit Power $P_m$ (W)& 40 & 1 \\ 
			\hline Intensity $\lambda_m$ ($\text{BS}$/km$^{2}$)  & $1$ & (see figures) \\ \hline  Pathloss Exponent &  \multicolumn{2}{c|}{4}  \\  \hline 
			$H_{m,i}$, $G_j$, $H_{m,i}$, $G_j$ & \multicolumn{2}{c|}{$\sim\mathrm{exp}(1)$}\\ \hline
			User Intensity $\mu$ (Users/km$^2$) & \multicolumn{2}{c|}{$500$}\\ \hline
			Transmit Power of Users $Q$ (mW) & \multicolumn{2}{c|}{$100$}\\ \hline
			Self-Interference Suppression Factor $\epsilon_*$ & \multicolumn{2}{c|}{$10^{-5}$}\\ \hline
			Self-Interference Suppression Factor $\epsilon_0$ & \multicolumn{2}{c|}{$10^{-8}$}\\ \hline
		\end{tabular}
	\end{center}
\end{table}

\begin{figure}[!t]
	\centering
	\includegraphics[width=6.8in, height=3.0in]{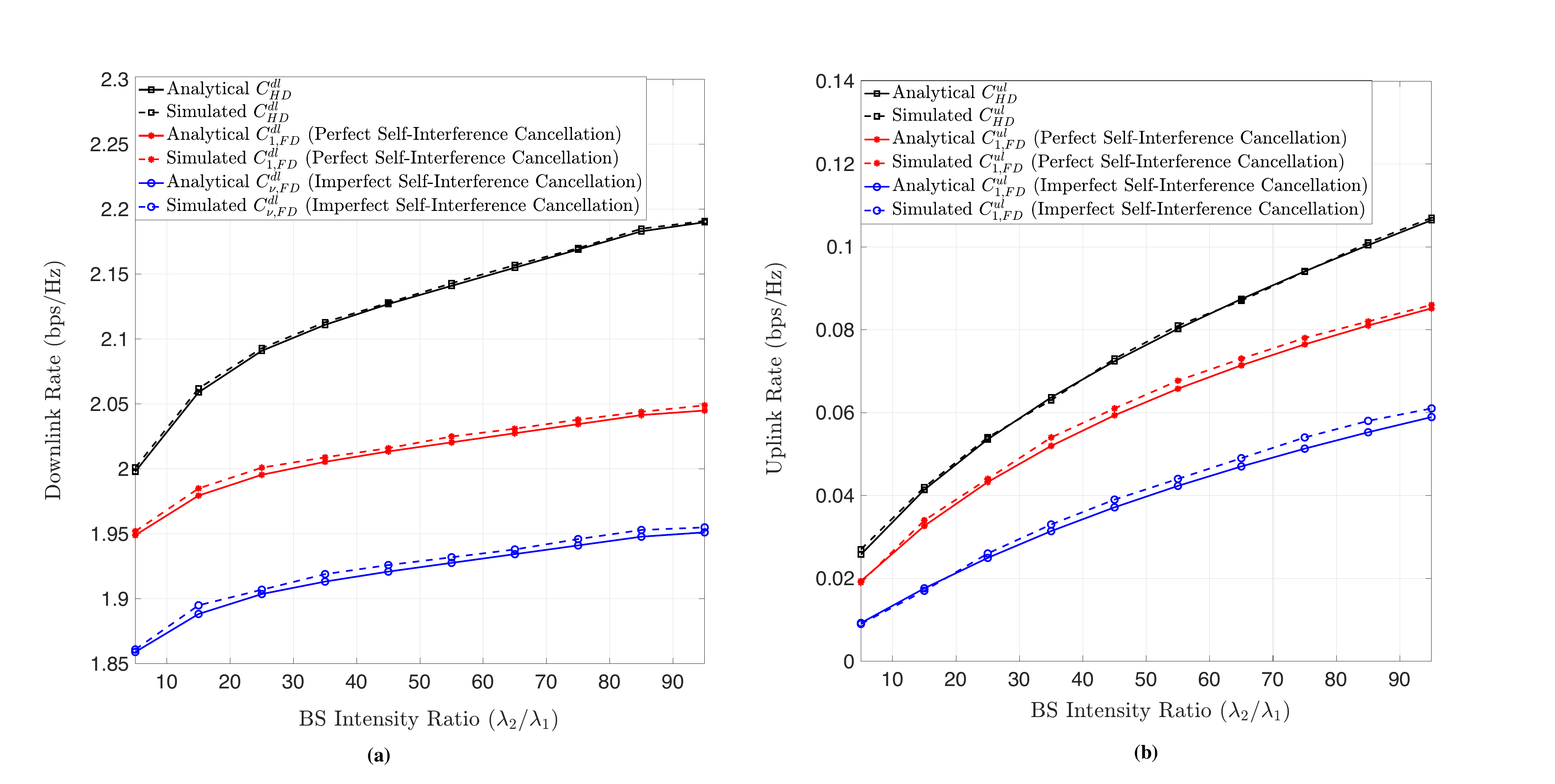}
	\caption {Simulation results of the achievable downlink and uplink rates: (a) Downlink rate, (b) Uplink rate. Note that $\nu=0$ corresponds to the case that all users are HD whereas $\nu=1$ corresponds to the case that all users are FD.
	}
	\label{Fig:DownUplink_rate}
\end{figure}

In the following simulation, we consider a two-tier HetNet.  For FD transmission, the MDROA schemes are adopted for decoupled user association, i.e., we have $\Psi^{\dl}_{m,i}(x)=P_mx^{-\alpha}$ and  $\Psi^{\ul}_{m,i}(x)=Qx^{-\alpha}$. All the network parameters used for simulation are shown in Table I. We first present the numerical results of the downlink and uplink rates in Fig. \ref{Fig:DownUplink_rate} in order to demonstrate whether or not the lower bounds on $C^{\dl}_{\nu,FD}$ and $C^{\ul}_{\nu,FD}$ in Proposition \ref{Prop:FDLinkrate} are really very tight and accurate. Indeed, all simulated results in Fig. \ref{Fig:DownUplink_rate} are slightly lower than their corresponding analytical upper bound results so that our previous analyses on the link rates are fairly correct and accurate. As shown in Fig. \ref{Fig:DownUplink_rate}, without a doubt the highest downlink and uplink rates are achieved by using HD transmission because there is no self-interference and FD interference at the receiver side. Also, we observe that self-interference needs to be suppressed as much as possible and otherwise it seriously undermines the SIR quality at the receiver side. 

\begin{figure}[!t]
	\centering
	\includegraphics[width=6.75in, height=2.85in]{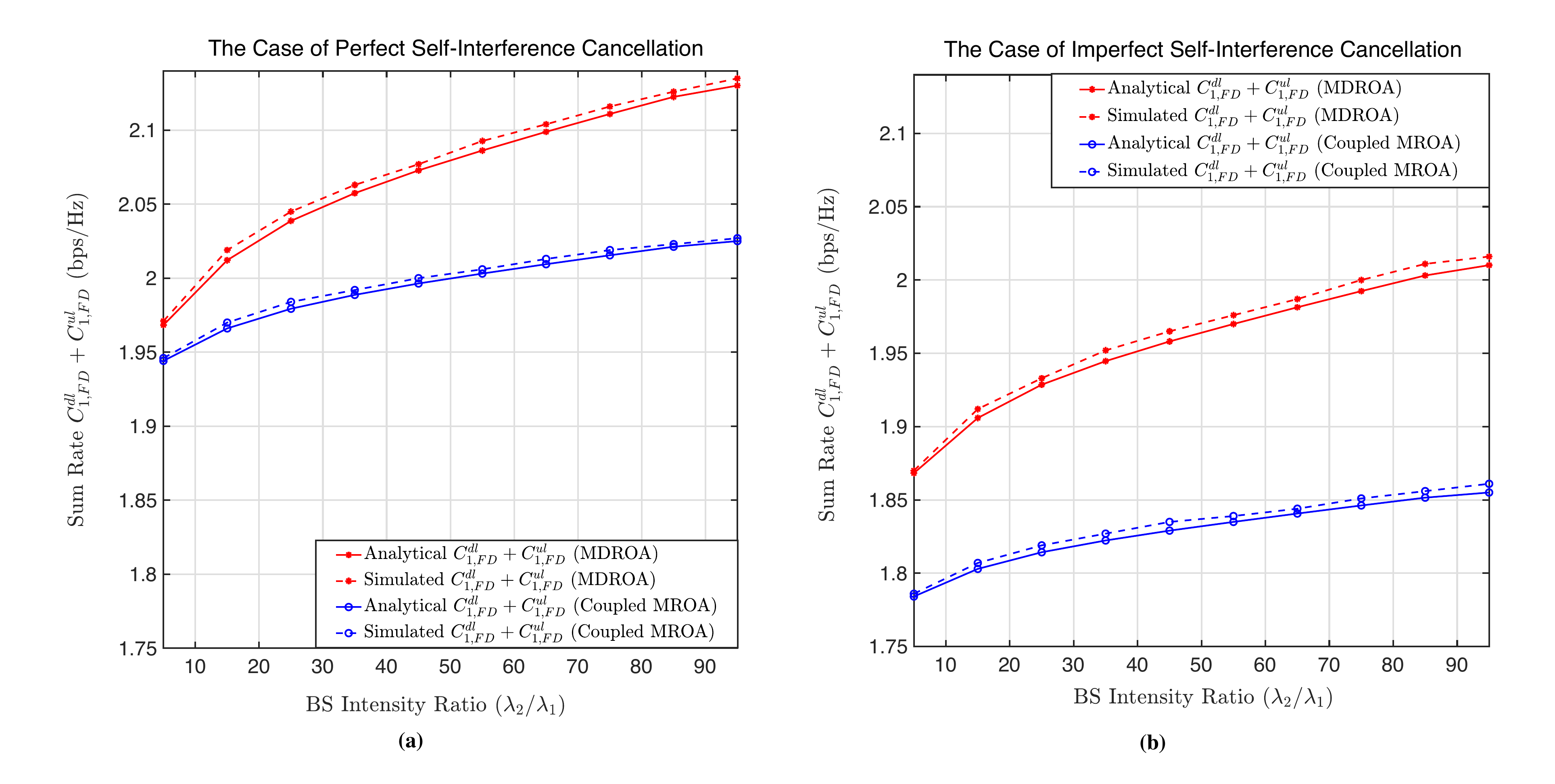}
	\caption{Simulation results of the sum of the downlink and uplink rates with the MDROA and coupled MROA scheme: (a) The case of perfect self-interference cancellation (b) The case of imperfect self-interference cancellation.}
	\label{Fig:DecoupledrateGain}
\end{figure}

  To evaluate how decoupled user association and FD transmission jointly influence the total rate of an FD user, let us define the sum rate of of an FD link between an FD user and its tagged BS as follows:
 \begin{align}\label{Eqn:SumThrooughputFD}
 C_{\nu,FD}\defn C^{\dl}_{\nu, FD}+C^{\ul}_{\nu, FD}.
 \end{align}
Note that $C^{\dl}_{\nu,FD}$ and $C^{\ul}_{\nu,HD}$ become $C^{\dl}_{1,FD}$ and $C^{\ul}_{1,FD}$ respectively provided all users in the HetNet are FD users (i.e., $\nu=1$)\footnote{For the case of $\nu=0$, this case means there is no FD traffic in the network, i.e., all users are HD users. For the case of $\nu\in(0,1)$, this case indicates that $\nu\%$ of the users in the network are FD users and the rest of the users in the network are HD users.}. The simulation results of the sum rate $C_{\nu,FD}$ for $\nu=1$ are shown in Fig. \ref{Fig:DecoupledrateGain} where there are two user association schemes simulated -- One is MDROA that is the decoupled user association scheme in \eqref{Eqn:UserAssFunROAMod} and the other is coupled MROA that is the ``modified" ROA scheme having the same user association function for downlink and uplink and using the mean channel gain of a tier as the association bias of the tier. Obviously, we can see that MDROA makes users achieve a much higher sum rate than coupled MROA in the both cases of perfect and imperfect self-interference cancellation. In addition, note that MDROA has an increasing rate gain much higher than coupled MROA as more and more BSs are deployed and this demonstrates that MDROA exploits the BS diversity induced by decoupled user association.

\section{Rate Regions and Opportunistic FD Scheduling}\label{Sec:RateRegion}

In Section \ref{SubSec:Simulationsrate}, we have numerically validated the correctness and accuracy of the tight lower bounds on link rates $C^{\dl}_{\nu, FD}$ and $C^{\ul}_{\nu, FD}$ found in Section \ref{SubSec:rateAnalysis}. In this section, our focus is on thoroughly exploiting the fundamental interplays between these two link rates. We will first characterize the achievable rate regions of the HD and FD users that can indicate how to optimally adopt the FD and HD modes in order to help users maintain high rate in different uplink and downlink traffic patterns. Then we will study how to scheduling bidirectional traffic in order to maximize the sum rate of a user.  

\subsection{Analysis of Achievable Rate Regions}
To start with the analysis of the achievable rate regions of users, first consider the scenario in which FD transmission is not allowed in the HetNet so that the downlink rate of an HD link is characterized by $C^{\dl}_{\nu, FD}$ in \eqref{Eqn:LinkrateNoSelfInt} with $\nu=\epsilon_0=0$, which is
\begin{align}\label{Eqn:PureHDrate}
C^{\dl}_{HD}\defn C^{\dl}_{\nu, FD}\big|_{\nu=\epsilon_0=0} \gtrapprox \sum_{m=1}^{M}\vartheta^{\dl}_m\int_{0^+}^{\infty}\frac{\left[1-\mathcal{L}_{H_m/\psi^{\dl}_m}(s)\right]}{s\left[\sum_{k=1}^{M}\vartheta^{\dl}_k\rho^{\dl}_k\Xi_1\left(Z,1,\frac{sP_kH_k}{P_m\psi^{\dl}_k}\right)+1\right]}\dif s
\end{align}
and it is the maximum achievable downlink rate for all HD users.  Similarly, in the uplink case, the maximum achievable uplink rate for all users in the HD mode, can be found as
\begin{align}
C^{\ul}_{HD}\defn C^{\ul}_{\nu, FD}\big|_{\nu=\epsilon_*=0}\gtrapprox \sum_{m=1}^{M}\vartheta^{\ul}_m\int_{0^+}^{\infty}\frac{\left[1-\mathcal{L}_{H_m/\psi^{\ul}_m}(s)\right]}{s\left\{\Gamma\left(1-\frac{2}{\alpha}\right)s^{\frac{2}{\alpha}}\sum_{k=1}^{M}\vartheta_k^{\ul}\rho^{\ul}_k\mathbb{E}\left[\left(\frac{P_kH_k}{Q\psi^{\ul}_k}\right)^{\frac{2}{\alpha}}\right]+1\right\}}\dif s.
\end{align}

\begin{figure}
	\centering
	\includegraphics[width=6.5in, height=2.5in]{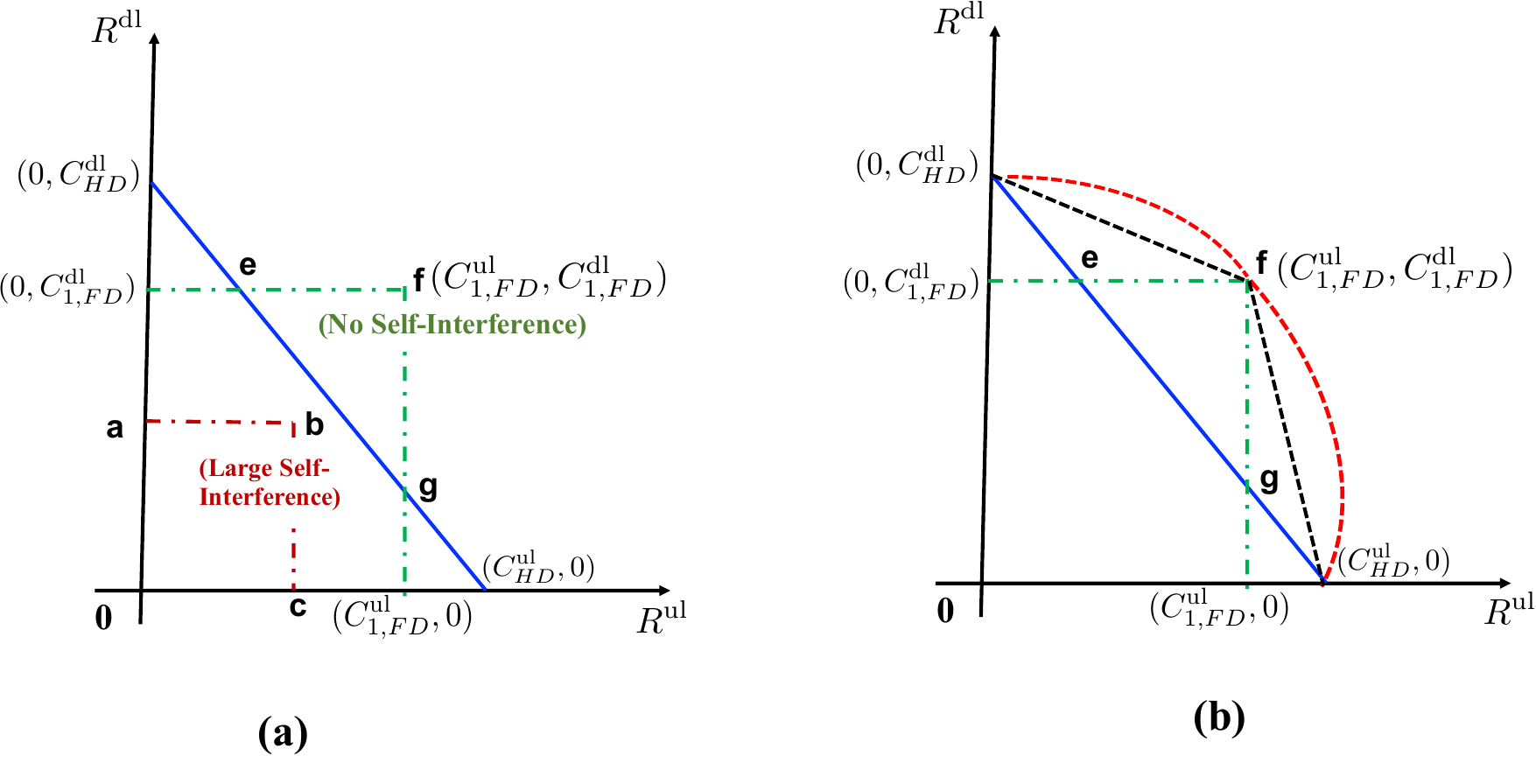}
	\caption{The rate regions of an FD user and an HD user where the horizontal axis denotes the uplink rate and the vertical axis represents the downlink rate: (a)  $\mathcal{R}_{HD}$ is the triangle with vertices \textbf{0}, $(0,C^{\dl}_{HD})$ and $(C^{\ul}_{HD},0)$,  and $\mathcal{R}_{FD}$ is the rectangle with vertices \textbf{0}, $(0,C^{\dl}_{1,FD})$, \textbf{f} and $(C^{\ul}_{1,FD},0)$. (b) The quadrilateral with vertices \textbf{0}, $(0,C^{\dl}_{HD})$, \textbf{f} and $(C^{\ul}_{HD},0)$, denoted by $\mathcal{R}_{\inf}$, is the convex hull of $\mathcal{R}_{HD}$ and $\mathcal{R}_{FD}$. The region enclosed by Line \textbf{f}-$(0,C^{\dl}_{HD})$, Line \textbf{f}-$(C^{\ul}_{HD},0)$ and Red Dash Line $(0,C^{\dl}_{HD})$-$(C^{\ul}_{HD},0)$, denoted by $\mathcal{R}_{\sup}$, is the maximum achievable rate region.}
	\label{Fig:rateregion}
\end{figure}

According to the above definitions of the link rates, we can characterize the achievable rate regions of an FD link.  By referring to $C_{\nu,FD}$ defined in \eqref{Eqn:SumThrooughputFD}, we define the rate region $\mathcal{R}_{FD}$ of an FD link for $\nu=1$ as follows:
\begin{align}
\mathcal{R}_{FD}\defn\left\{(R^{\ul},R^{\dl})\in\mathbb{R}^2_{+}: R^{\ul}\leq C^{\ul}_{1,FD}, R^{\dl}\leq C^{\dl}_{1,FD}\right\}.
\end{align} 
Namely, this region is the rate region of an FD  link when all users in the HetNet are FD. Whereas the rate region $\mathcal{R}_{HD}$ of an HD link can be defined as
\begin{align}
\mathcal{R}_{HD}\defn\left\{(R^{\ul},R^{\dl})\in\mathbb{R}^2_+: R^{\ul}\leq\theta C^{\ul}_{HD}, R^{\dl}\leq(1-\theta) C^{\dl}_{HD},\theta\in[0,1] \right\},
\end{align}
where $\theta$ is the time-sharing parameter for the HD transmission between downlink and uplink. The rate regions, $\mathcal{R}_{FD}$ and $\mathcal{R}_{HD}$, can be schematically demonstrated in Fig. \ref{Fig:rateregion} where the horizontal axis denotes the uplink rate whereas the vertical axis represents the downlink rate.
As shown in Fig. \ref{Fig:rateregion}, Region $\mathcal{R}_{FD}$ is essentially the rectangle with vertices \textbf{0}, $(0,C^{\dl}_{1, FD})$, \textbf{f} and $(C^{\ul}_{1, FD},0)$ for the case of no self-interferences since every point in this rectangle is achievable by FD transmission and perfect self-interference cancellation. The triangle with vertices \textbf{0}, $(0,C^{\dl}_{HD})$ and $(C^{\ul}_{HD},0)$ is region $\mathcal{R}_{HD}$ because the points on Line $(0,C^{\dl}_{HD})$-$(C^{\ul}_{HD},0)$ can be achieved by altering the time-sharing parameter $\theta$ between 0 and 1. Note that we have $C^{\dl}_{1, FD}<C^{\dl}_{HD}$ and $C^{\ul}_{1, FD}<C^{\ul}_{HD}$ due to the FD interferences. When the self-interferences cannot be canceled and fairly large, region $\mathcal{R}_{FD}$ would shrink to the much smaller rectangle with vertices \textbf{0}, \textbf{a}, \textbf{b} and \textbf{c}, which demonstrates that using FD does not outperform HD in terms of link rates at the presence of large self-interferences.

As can be seen in Fig. \ref{Fig:rateregion} (a), we realize that $\mathcal{R}_{HD}$ is not enclosed by $\mathcal{R}_{FD}$, i.e., $\mathcal{R}_{HD}\nsubseteq \mathcal{R}_{FD}$, which essentially clarifies that \textit{using FD does not always achieve a larger rate region than using HD}. Considering the triangular region  with vertices $(0,C^{\dl}_{1, FD})$, \textbf{e}, and $(0,C^{\dl}_{HD})$, for instance, any points in this region cannot be achieved by using FD and its sum rate could be higher than $C_{1,FD}=C^{\dl}_{1,FD}+C^{\ul}_{1,HD}$. On the contrary, using HD cannot achieve any points in the triangle with vertices \textbf{e}, \textbf{f} and \textbf{g}.  A larger achievable rate region that encloses $\mathcal{R}_{HD}$ and $\mathcal{R}_{FD}$ indeed exists, as stated in the following proposition.
\begin{proposition}\label{Prop:rateRegion}
According to Fig. \ref{Fig:rateregion} (b), the convex region enclosed by Line \textbf{0}-$(0,C^{\dl}_{HD})$, Red Dash Line $(0,C^{\dl}_{HD})$-\textbf{f}-$(C^{\ul}_{HD},0)$ and Line $(C^{\ul}_{HD},0)$-\textbf{0}, denoted by $\mathcal{R}_{\sup}$, is the maximum achievable rate region of a user in the HetNet with decoupled user association. 
\end{proposition} 
\begin{IEEEproof}
See Appendix \ref{App:ProofrateRegion}.
\end{IEEEproof}

The rate regions in Fig. \ref{Fig:rateregion} and Proposition \ref{Prop:rateRegion} reveal some crucial implications worth addressing as follows:
\begin{itemize}
	\item Whenever large self-interference exists at either a user or a BS or both, we should avoid using the FD mode because in this situation using the FD transmission may not improve or even reduce the link rates, as illustrated in Fig. \ref{Fig:rateregion}(a).  
	\item The uplink and downlink rate pair at point \textbf{f} is the case of $\nu=1$, and for this case all users in the HetNet are FD. When $\nu$ reduces and approaches to zero, point \textbf{f} will move towards to Line $(0,C^{\dl}_{HD})$-$(C^{\ul}_{HD},0)$, point $(0,C^{\dl}_{1,FD})$ will move up to point $(0,C^{\dl}_{HD})$ and point $(C^{\ul}_{1,FD},0)$ will move right to point $(C^{\ul}_{HD},0)$. Namely, $\mathcal{R}_{FD}$ will gradually become $\mathcal{R}_{HD}$ as $\nu$ decreases from unity to zero. Accordingly, the bidirectional traffic pattern characterized by $\nu$ between a user and its tagged BS intrinsically dominates the rate region so that properly controlling the bidirectional traffic helps us improve the sum rate of a user. 
	\item In order to achieve the largest rate region $\mathcal{R}_{\sup}$, it is necessary to do time-sharing between points \textbf{f} and $(0,C^{\dl}_{HD})$ and time-sharing between points \textbf{f} and $(C^{\ul}_{HD},0)$, i.e., exclusively using FD or HD cannot achieve the maximum sum rate and it is necessary to schedule the downlink and uplink traffic by adopting the FD and HD modes appropriately and alternatively.
\end{itemize}

To sum up, using FD all the time in the HetNet does not always achieve higher rate than using HD so that we need to schedule the bidirectional traffic between downlink and uplink by using the FD and HD modes properly in order to maximize the sum rate of the bidirectional traffic. Some numerical results will be given in Section \ref{SubSec:NumResultScheduling} to illustrate this conclusion. 

\subsection{Opportunistic FD Scheduling Algorithms and Their Stability}
In this subsection, we would like to propose traffic scheduling algorithms to achieve the largest rate region $\mathcal{R}_{\sup}$. Consider a BS and its serving user both have a buffer of infinite size storing their packages. According to Fig. \ref{Fig:rateregion} (b), we propose the following opportunistic FD scheduling algorithms for achieving the points in $\mathcal{R}_{\sup}$:
\begin{algorithm}[Downlink Opportunistic FD Scheduling]\label{Alg:Downlink}
 Consider that downlink traffic is more than uplink traffic: (i) if both uplink and downlink queues are not empty, use the FD mode for bidirectional transmission; (ii) if only the uplink queue is empty, use the HD mode for downlink transmission; (iii) no transmission is scheduled if only the downlink queue is empty. Such a scheduling algorithm achieves the rate point on the Red Dash Line between points \textbf{f} and $(0,C^{\dl}_{HD})$.  
\end{algorithm} 
\begin{algorithm}[Uplink Opportunistic FD Scheduling]\label{Alg:Uplink}
 Consider that uplink traffic is more than downlink traffic: (i) if both uplink and downlink queues are not empty, use the FD mode for bidirectional transmission; (ii) if only the downlink queue is empty, use the HD mode for uplink transmission; (iii) no transmission is scheduled if only the uplink queue is empty. Such a scheduling algorithm achieves the rate point on the Red Dash Line between points \textbf{f} and $(C^{\ul}_{HD},0)$.
\end{algorithm} 

\noindent Note that the condition of achieving the maximum of sum rate $C_{\nu,FD}$ on the Red Dash Line is given by
\begin{align}
\frac{\dif C_{\nu,FD}}{\dif \nu}=0\Rightarrow \frac{\dif C^{\ul}_{\nu,FD}/\dif \nu}{\dif C^{\dl}_{\nu,FD}/\dif\nu} = \frac{\dif C^{\ul}_{\nu,FD}}{\dif C^{\dl}_{\nu,FD}}\bigg|_{\nu=\nu_*} = -1,
\end{align}
where $\nu_*$ is the optimal value (traffic pattern) of satisfying this condition. Therefore, \textit{the above two scheduling algorithms are rate-optimal while they are performed with traffic pattern $\nu=\nu_*$}.  

For any Poisson packet arrival processes with a (bit-arrival) rate pair within $\mathcal{R}_{\sup}$, the salient feature of the scheduling algorithms proposed above is that they do not require any arrival rate information of the downlink and uplink queues to achieve the stability of these two queues, as stated in the following proposition.
\begin{proposition}\label{Prop:QueueingStability}
Suppose all packets in the HetNet have the same size of $\ell$ and they arrive their own buffers according to a Poisson process. If packets arrive at a BS with rate $\eta^{\dl}$ and packets arrive at a user with rate $\eta^{\ul}$, Algorithms \ref{Alg:Downlink} and \ref{Alg:Uplink} stabilize the uplink and downlink queues as long as the (bit-arrival) rate pair $(\eta^{\ul}\ell,\eta^{\dl}\ell)$ is within $\mathcal{R}_{\sup}$. 
\end{proposition}
\begin{IEEEproof}
See Appendix \ref{App:ProofQueueingStability}.
\end{IEEEproof}

\noindent Based on Proposition \ref{Prop:QueueingStability}, we can ensure that Algorithms \ref{Alg:Downlink} and \ref{Alg:Uplink} are able to maximize the sum of the downlink and uplink rates as well as stabilize the downlink and uplink queues.

\subsection{Numerical Results}\label{SubSec:NumResultScheduling}
\begin{figure}
	\centering
	\includegraphics[width=6.85in, height=2.85in]{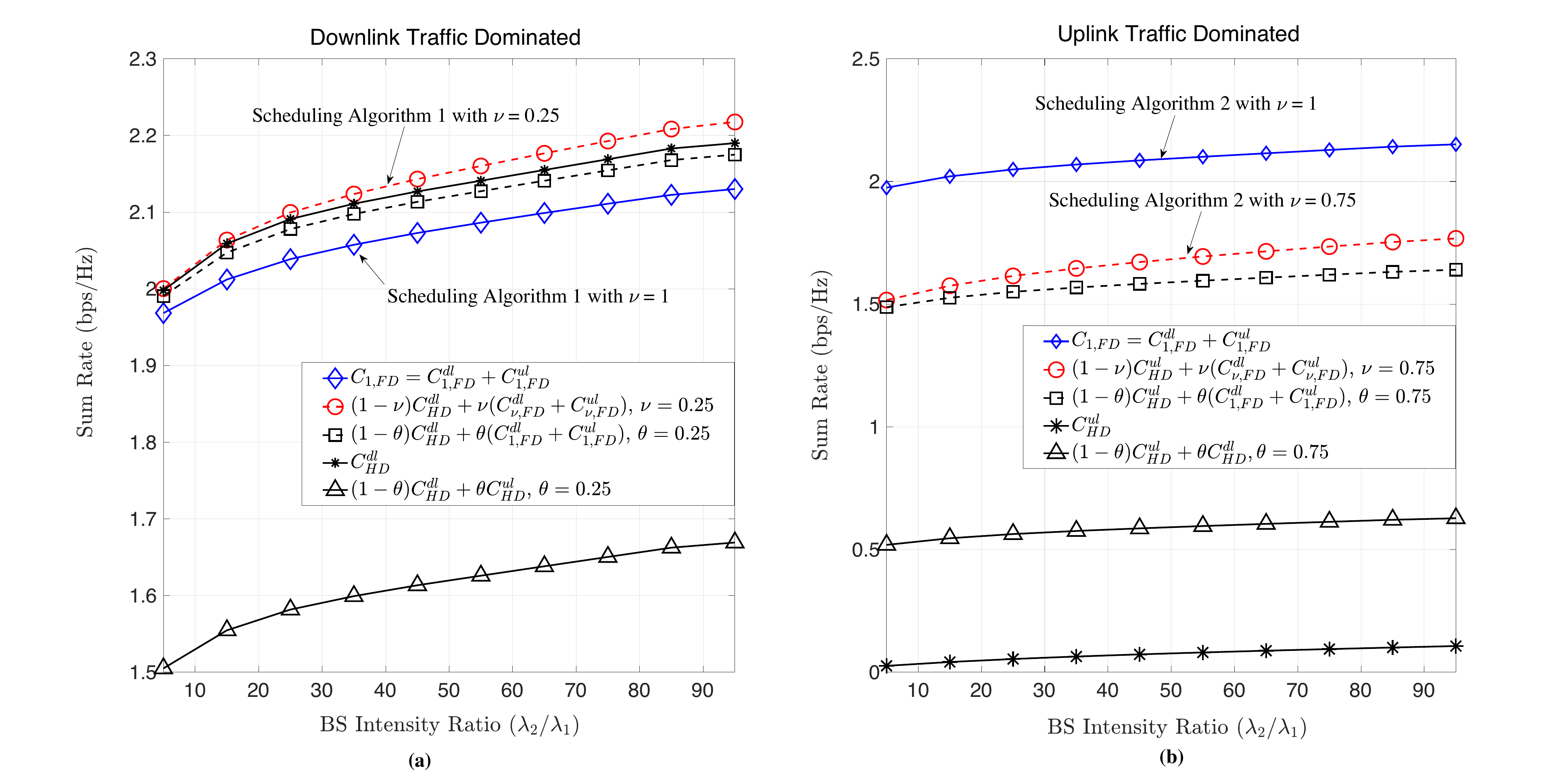}
	\caption{Simulation results of the sum rate of a user: (a) The case of the HetNet with dominated downlink traffic, (b) The case of the HetNet with dominated uplink traffic}
	\label{Fig:rateScheduling}
\end{figure}

To illustrate that Scheduling Algorithms \ref{Alg:Downlink} and \ref{Alg:Uplink} indeed improve the sum rate of a user, some of the numerical results with implementing Scheduling Algorithms \ref{Alg:Downlink} and \ref{Alg:Uplink} are shown in Fig. \ref{Fig:rateScheduling} by using the same simulation settings and network parameters in Section \ref{SubSec:Simulationsrate}. In Fig. \ref{Fig:rateScheduling} (a), we consider the circumstance that there is much more downlink traffic than uplink traffic in the HetNet. For the pure HD case, 75\% of the total transmission time is used for downlink traffic and 25\% of the total transmission time are occupied by uplink traffic and thus the sum rate is $0.75C^{\dl}_{HD}+0.25C^{\ul}_{HD}$. For the downlink opportunistic FD scheduling, we consider $\nu$ as the time-sharing parameter, i.e., we have $\nu$ fraction of the total transmission time used for FD transmission and ($1-\nu$) fraction of the total transmission time used for HD downlink transmission so that the the sum rate is $(1-\nu)C^{\dl}_{HD}+\nu C_{\nu,FD}$. As can be observed in Fig. \ref{Fig:rateScheduling} (a), using Scheduling Algorithm \ref{Alg:Downlink} with $\nu=0.25$ attains the highest rate among all sum rates, which indicates that Scheduling Algorithm \ref{Alg:Downlink} indeed achieves the largest rate region $\mathcal{R}_{\sup}$ as shown in Fig. \ref{Fig:rateregion} (b). The simulation result for the circumstance that the uplink traffic largely dominates the downlink traffic is shown in Fig. \ref{Fig:rateScheduling} (b) and we also can see the similar phenomenon that the sum rate achieved by Scheduling Algorithm \ref{Alg:Uplink} is superior to the sum rates achieved by purely using HD and FD. Finally, the simulation results in Fig. \ref{Fig:rateScheduling} reveal a key point; that is, we can optimize $\nu$ and use it to perform Scheduling Algorithm \ref{Alg:Downlink} or \ref{Alg:Uplink} so as to achieve the maximum rate in $\mathcal{R}_{\sup}$.

\section{Conclusions}\label{Sec:Conclusion}
Full-duplex transmission is a promising transmission technique that could double the rate for a P2P link. However, the rate performance of FD is very unclear in a large-scale network where there are more interferences induced by FD transmission if compared with HD transmission. To comprehensively study the uplink and downlink rates of an FD link between a user and its tagged in a large-scale cellular network, we propose a HetNet model in which users/BSs can be HD or FD depending whether they have FD traffic at the same time. To characterize the downlink and uplink rates of a FD link, we find the DROA scheme that helps us characterize the full-duplex SIR model and the uplink and downlink rates of an FD link. The tight lower bounds on $C^{\dl}_{\nu,FD}$ and $C^{\ul}_{\nu,FD}$ are found in a very general form and their tightness and accuracy are verified numerically. We use them not only to show that using FD transmission in a HetNet may not achieve the largest sum rate of an FD link in the uplink and downlink, but also to delineate the rate regions of an FD link in the uplink and downlink contexts. These rate regions clearly manifest that the maximum rate region of an FD link indeed exists and scheduling the uplink and downlink traffic by opportunistically using the FD and HD modes can maximize the sum of the uplink and downlink rate of an FD link. As a result, two opportunistic FD scheduling algorithms are proposed and they are theoretically and numerically shown to achieve the largest rate region with queuing stability.

\appendix[Proofs of Lemmas and Propositions]

\subsection{Proof of Lemma \ref{Lem:LapTransShotNoise}}\label{App:ProofLapTransShotNoise}
Since $Y_{i}$ is the $i$th nearest point in $\mathcal{Y}$ to the origin, we know $\|Y_i\|^2$ is the sum of $i$ i.i.d. exponential RVs with probability density function (pdf) $f_{Y_1}(y)=\pi\lambda_{\mathcal{Y}}e^{-\pi\lambda_{\mathcal{Y}}y}$. Then $\mathcal{L}_{\mathfrak{I}_n}(s)$ can be found as follows
\begin{align*}
\mathcal{L}_{\mathfrak{I}_n}(s) &= \mathbb{E}\left[\exp\left(-s\sum_{i:Y_i\in\mathcal{Y}}W_{n+i} \xi\left(\|Y_{n+i}\|^2\right)\right)\right]=\mathbb{E}\left[\exp\left(-s\sum_{i:Y_i\in\mathcal{Y}}W_{n+i} \xi\left(\|Y_n\|^2+\|Y_{i}\|^2\right)\right)\right]
\end{align*}
because  $\|Y_n\|^2$ and $\|Y_i\|^2$ are independent so that we have the identity $\|Y_{n+i}\|^2=\|Y_n\|^2+\|Y_i\|^2$ (the proof of this identity can be found in \cite{CHLLCW16,MH12}). By conditioning on $\|Y_n\|^2$ and using the probability generation functional (PGFL) of a homogeneous PPP \cite{DSWKJM13}, we can get
\begin{align*}
\mathbb{E}\left[\exp\left(-s\sum_{i:Y_i\in\mathcal{Y}}W_{n+i} \xi\left(\|Y_n\|^2+\|Y_{i}\|^2\right)\right)\bigg|\|Y_n\|^2\right]= e^{-\pi\lambda_{\mathcal{Y}}\int_{0}^{\infty}\mathbb{E}\left[1-e^{-sW\xi(\|Y_n\|^2+r)}\right]r\dif r }
\end{align*}
and letting $\mathsf{E}\sim\exp(1)$ be an exponential RV with unit mean and variance yields the following
\begin{align*}
\mathbb{P}\left[Z\leq sW\xi(\|Y_n\|^2+r)\right]=\mathbb{P}\left[\xi^{-1}\left(\frac{\mathsf{E}}{sW}\right)\geq \|Y_n\|^2+r\right]=\mathbb{E}\left[1-e^{-sW\xi(\|Y_n\|^2+r)}\right].
\end{align*}
Thus, letting $u\defn \|Y_n\|^2 +r$ and considering $n\neq 0$ yield
\begin{align*}
&\mathcal{L}_{\mathfrak{I}_n}(s) = \mathbb{E}_{\|Y_n\|^2}\left[\exp\left\{-\pi\lambda_{\mathcal{Y}} \int_{\|Y_n\|^2}^{\infty}\mathbb{P}\left[\xi^{-1}\left(\frac{\mathsf{E}}{sW}\right)\geq u\right] \dif u\right\}\right]\\
&=\mathbb{E}_{\|Y_n\|^2}\left[\exp\left\{-\pi\lambda_{\mathcal{Y}} \left(\mathbb{E}\left[\xi^{-1}\left(\frac{\mathsf{E}}{sW}\right)\right]-\int^{\|Y_n\|^2}_{0}\left(1-\mathbb{P}\left[\xi^{-1}\left(\frac{\mathsf{E}}{sW}\right)\leq u\right]\right) \dif u\right)\right\}\right]\\
&=\mathbb{E}_{\|Y_n\|^2}\left[\exp\left\{-\pi\lambda_{\mathcal{Y}} \left(\mathbb{E}\left[\xi^{-1}\left(\frac{\mathsf{E}}{sW}\right)\right]-\|Y_n\|^2+\int^{\|Y_n\|^2}_{0}\mathcal{L}_{W}(s\xi(u)) \dif u\right)\right\}\right]
\end{align*}
\begin{align*}
&=\mathbb{E}_{\|Y_n\|^2}\left[\exp\left\{-\pi\lambda_{\mathcal{Y}} \left(\mathbb{E}\left[\xi^{-1}\left(\frac{\mathsf{E}}{sW}\right)\right]+\|Y_n\|^2\left[\int^{1}_{0}\mathcal{L}_{sW}(\xi(v\|Y_n\|^2)) \dif v-1\right]\right)\right\}\right]\\
&=\int_{0}^{\infty} \exp\left[-\pi\lambda_{\mathcal{Y}}\Xi_{1}\left(\mathsf{E},y,sW\right)\right]f_{\|Y_n\|^2}(y)\dif y.
\end{align*}
Whereas for $n=0$, we can get $\mathcal{L}_{\mathfrak{I}_0}(s)=\exp\left[-\pi\lambda_{\mathcal{Y}} \Xi_{0}\left(\frac{\mathsf{E}}{W},0\right)\right]$.  Therefore, the result in \eqref{Eqn:LapTransShotNoise} is acquired by the fact that $\|Y_n\|^2$ is a sum of $n$ i.i.d. exponential RVs with mean $1/\pi\lambda_{\mathcal{Y}}$ and it is essentially a Gamma RV with shape parameter $n$ and rate parameter $\pi\lambda_{\mathcal{Y}}$.

\subsection{Proof of Lemma \ref{Lem:DROA}}\label{App:ProofDROA}
First of all, consider the downlink case. The SIR $\gamma^{\dl}_{m,i}(\|X_{m,i}\|)$ can be written as
\begin{align*}
\gamma^{\dl}_{m,i} (\|X_{m,i}\|)= \frac{P_mH_{m,i}\|X_{m,i}\|^{-\alpha}}{I'_0-P_mH_{m,i}\|X_{m,i}\|^{-\alpha}}=\left(\frac{I'_0\|X_{m,i}\|^{\alpha}}{P_mH_{m,i}}-1\right)^{-1},
\end{align*}
where $I'_0$ denotes the total signal power received by the typical user and this follows that
\begin{align*}
X^{\dl}_*&=\arg\sup_{m,i:X_{m,i}\in\mathcal{X}}\left\{\log\left(1+\gamma_{m,i}[\|X_{m,i}\|]\right)\right\}\stackrel{(a)}{=}\arg\sup_{m,i:X_{m,i}\in\mathcal{X}} \{\gamma_{m,i}(\|X_{m,i}\|)\}\\
&=\arg\sup_{m,i:X_{m,i}\in\mathcal{X}} \left\{\left(\frac{I'_0\|X_{m,i}\|^{\alpha}}{P_mH_{m,i}}-1\right)^{-1}\right\}\stackrel{(b)}{=}\arg\sup_{m,i:X_{m,i}\in\mathcal{X}} \left\{\frac{P_mH_{m,i}}{\|X_{m,i}\|^{\alpha}}\right\},
\end{align*}
where $(a)$ follows from the fact that $\log(1+x)$ and $x$ give rise to the same association result and $(b)$ is due to the fact that random variable $I'_0$ is the same for different BSs and does not affect the user association result and thus $I'_0$ can be removed. Hence, using $\Psi_{m,i}(\|X_{m,i}\|)=P_mH_{m,i}\|X_{m,i}\|^{-\alpha}$ can make the typical user associate with a BS that provides the maximum downlink rate to it.

Now consider the uplink case. The SIR $\gamma^{\ul}_{m,i}(\|X_{m,i}\|)$ can be expressed as
\begin{align*}
\gamma^{\ul}_{m,i}(\|X_{m,i}\|) = \frac{Q\breve{H}\|X_{m,i}\|^{-\alpha}}{I'_*-Q\breve{H}\|X_{m,i}\|^{-\alpha}}=\left(\frac{I'_*\|X_{m,i}\|^{\alpha}}{Q\breve{H}_{m,i}}-1\right)^{-1}.
\end{align*}
According to \eqref{Eqn:RateOptUserAssFun}, the BS that provides the maximum uplink rate to the typical user is written as
\begin{align*}
X_*^{\ul} &= \arg\sup_{m,i:X_{m,i}\in\mathcal{X}}\log\left(1+\gamma^{\ul}_{m,i}(\|X_{m,i}\|) \right)=\arg\inf_{m,i:X_{m,i}\in\mathcal{X}}\left\{\frac{I'_*\|X_{m,i}\|^{\alpha}}{Q\breve{H}_{m,i}}\right\}\\
&= \arg\sup_{m,i:X_{m,i}\in\mathcal{X}}\left\{\frac{Q\breve{H}_{m,i}}{I'_*\|X_{m,i}\|^{\alpha}}\right\}\stackrel{(c)}{=}\arg\sup_{m,i:X_{m,i}\in\mathcal{X}}\left\{\frac{\breve{H}_{m,i}}{\|X_{m,i}\|^{\alpha}}\right\},
\end{align*}
where $(c)$ follows from that fact that $Q/I'_*$ is i.i.d. at different BSs and removing it does not affect the result of user association based on the Slivnyak theorem \cite{DSWKJM13} and Theorem 1 in \cite{CHLLCW16}. Thus, letting $\Psi^{\ul}_{m,i}(\|X_{m,i}\|)=\breve{H}_{m,i}\|X_{m,i}\|^{-\alpha}$ makes users associate with the uplink BS that is able to provide the maximum uplink rate to it. Therefore, the GUA scheme with $\Psi_{m,i}(\cdot)$ given in \eqref{Eqn:UserAssFunROA} is exactly the DROA scheme in \eqref{Eqn:RateOptUserAssFun}.

\subsection{Proof of Proposition \ref{Prop:FDLinkrate}}\label{App:ProofFDLinkrate}

Before proceeding the proof, we first need to introduce the integral identity of the Shannon transformation in Theorem 1 in our previous work \cite{CHLHCT1701} as follows: For a non-negative RV $Z$, its Shannon transform is defined as $\mathcal{S}_Z(h) \defn \mathbb{E}[\log (1+hZ)]$ with parameter $h>0$ and it has an integral identity given by
\begin{align}
\mathcal{S}_Z(h) = \int_{0}^{\infty} \frac{[1-\mathcal{L}_{h}(s)]}{s}\mathcal{L}_{Z^{-1}}(s) \dif s
\end{align}
if $h$ is a RV independent from $Z$ and its Laplace transform exists. Using this integral identity and $\gamma^{\dl}_0$ defined in \eqref{Eqn:DownlinkUserSIR}, we can rewrite $C^{\dl}_{\nu, FD}$ as
\begin{align*}
C^{\dl}_{\nu, FD} & =\mathbb{E}\left[\log\left(1+\frac{\gamma^{\dl}_0H_*}{H_*}\right)\right]=\sum_{m=1}^{M}\vartheta^{\dl}_m\mathbb{E}\left[\int_{0}^{\infty} \frac{1}{s}\left(1-e^{-s\frac{H_m}{\psi^{\dl}_m}}\right)\mathcal{L}_{\frac{H_m}{\psi^{\dl}_m\gamma^{\dl}_0}}(s)\dif s\right].
\end{align*}
The Laplace transform of $\frac{H_m}{\psi^{\dl}_m\gamma_0}$ can be explicitly written and further simplified as shown in the following:
\begin{align*}
\mathcal{L}_{\frac{H_m}{\psi^{\dl}_m\gamma_0}}(s)=&\mathbb{E}\left[\exp\left( \frac{ -s(I_0+\epsilon_0 Q)}{P_m\psi^{\dl}_m \|X^{\dl}_*\|^{-\alpha}}\right)\right]\\
\stackrel{(a)}{=}& \mathbb{E}\left[\exp\left(-\frac{s\|\widetilde{X}^{\dl}_*\|^{\alpha}}{P_m}\left(\sum_{k,i\in\widetilde{\mathcal{X}}\setminus \widetilde{X}^{\dl}_m}\frac{P_kV^{\dl}_{k,i}H_{k,i}}{\psi^{\dl}_{k,i}\|\widetilde{X}_{k,i}\|^{\alpha}}+I^{\dl}_{\mathcal{U}}+\epsilon_0 Q\right)\right)\right]\\
=& \mathbb{E}\left\{\exp\left(-\frac{s}{P_m}\sum_{k,i\in\widetilde{\mathcal{X}}\setminus X^{\dl}_*}\frac{P_kV^{\dl}_{k,i}H_{k,i}}{\psi^{\dl}_{k,i}(\|\widetilde{X}_{k,i}\|^2/\|\widetilde{X}_*^{\dl}\|^2)^{\frac{\alpha}{2}}}\right)\cdot\exp\left(-\frac{s\|\widetilde{X}^{\dl}_*\|^{\alpha}}{P_m}(I^{\dl}_{\mathcal{U}}+\epsilon_0 Q)\right)\right\},
\end{align*}
where $(a)$ follows by letting $\mathcal{X}\defn \mathcal{X}_{m}$ in which $\widetilde{\mathcal{X}}_m\defn\{\widetilde{X}_{m,i}\in\mathbb{R}^2: \widetilde{X}_{m,i}= (\psi^{\dl}_{m,i})^{-\frac{1}{\alpha}}X_{m,i} \}$ and set $\widetilde{\mathcal{X}}_m$ is a homogeneous PPP of intensity $\widetilde{\lambda}_m$ based on the conservation property in Theorem 1 in \cite{CHLLCW16}. Thus, set $\widetilde{\mathcal{X}}$ is also a homogeneous PPP of intensity $\widetilde{\lambda}^{\dl}=\sum_{m=1}^{M}\widetilde{\lambda}^{\dl}_m$ since all $\widetilde{\mathcal{X}}_m$'s are independent. Note that $\widetilde{X}^{\dl}_*$ is the nearest point in set $\widetilde{\mathcal{X}}$ to the typical user. In addition, although all $V^{\dl}_{k,i}$'s are not completely independent based on the results in \cite{CHLLCW1502}\cite{CHLLCW16}, the correlations among them are fairly weak in general.

According to Lemma \ref{Lem:LapTransShotNoise}, we can have the following result:
\begin{align*}
\mathbb{E}\left\{\exp\left[-\frac{s}{P_m}\sum_{k,i\in\widetilde{\mathcal{X}}\setminus \widetilde{X}^{\dl}_*}\frac{P_kV^{\dl}_{k,i}H_{k,i}}{\psi^{\dl}_{k,i}}\left(\frac{\|\widetilde{X}_{k,i}\|^2}{\|\widetilde{X}_*^{\dl}\|^2}\right)^{-\frac{\alpha}{2}}\right]\right\}=\mathcal{L}_{\mathfrak{I}_1}\left(\frac{s\|\widetilde{X}^{\dl}_*\|^{\alpha}}{P_m}\right),
\end{align*} 
where $\mathcal{L}_{\mathfrak{I}_1}(\cdot)$ is given by
\begin{align*}
\mathcal{L}_{\mathfrak{I}_1}\left(\frac{s\|\widetilde{X}^{\dl}_*\|^{\alpha}}{P_m}\right) = \prod_{k=1}^{M}\mathcal{L}_{\Xi_1\left(Z,\|\widetilde{X}^{\dl}_*\|^2,\frac{sP_kV^{\dl}_kH_k\|\widetilde{X}^{\dl}_*\|^{\alpha}}{P_m\psi^{\dl}_k}\right)}\left(\pi\widetilde{\lambda}_k\right) 
\end{align*}
and $\Xi_1(\cdot,\cdot,\cdot)$ for letting $\xi(x) = x^{-\alpha/2}$ and $\xi^{-1}(x)= x^{-2/\alpha}$ is given by
\begin{align*}
&\Xi_1\left(Z,\|\widetilde{X}^{\dl}_*\|^2,\frac{sP_kV^{\dl}_kH_k\|\widetilde{X}^{\dl}_*\|^{\alpha}}{P_m\psi^{\dl}_k}\right)= \Gamma\left(1-\frac{2}{\alpha}\right) \mathbb{E}\left[\|\widetilde{X}^{\dl}_*\|^2\left(\frac{sP_kV^{\dl}_kH_k}{P_m\psi^{\dl}_k}\right)^{\frac{2}{\alpha}}\right]+\|\widetilde{X}^{\dl}_*\|^2\\
&\left(\int_{0}^{1} \mathbb{E}\left[\exp\left(-\frac{sP_kV^{\dl}_kH_k}{\psi^{\dl}_kP_mv^{\alpha/2}}\right)\right]\dif v-1\right)\stackrel{(b)}{\gtrapprox}\rho^{\dl}_k\|\widetilde{X}^{\dl}_*\|^2\bigg[\Gamma\left(1-\frac{2}{\alpha}\right)\mathbb{E}\left[\left(\frac{sP_kH_k}{P_m\psi^{\dl}_k}\right)^{\frac{2}{\alpha}}\right]\\
&+\int_{0}^{1} \mathcal{L}_{\frac{sP_kH_k}{P_m\psi^{\dl}_k}}\left(v^{-\frac{\alpha}{2}}\right)\dif v-1\bigg]\stackrel{(c)}{=}\rho^{\dl}_k\|\widetilde{X}^{\dl}_*\|^2\Xi_1\left(\mathsf{E},1,\frac{sP_kH_k}{P_m\psi^{\dl}_k}\right),
\end{align*}
where $(b)$ follows from the fact that the correlations among all $V^{\dl}_{k,i}$'s are fairly weak and  assuming they are independent just slightly increases the interference from all BSs, and $(c)$ follows from the result in Lemma \ref{Lem:LapTransShotNoise} for $n=1$. Thus, we have
\begin{align}
\mathcal{L}_{\mathfrak{I}_1}\left(\frac{s\|\widetilde{X}^{\dl}_*\|^{\alpha}}{P_m}\right) =\mathbb{E}_{\|\widetilde{X}^{\dl}_*\|^2}\left[\exp\left(-\pi\|\widetilde{X}^{\dl}_*\|^2\sum_{k=1}^{M}\widetilde{\lambda}^{\dl}_k\rho^{\dl}_k\Xi_1\left(\mathsf{E},1,\frac{sP_kH_k}{P_m\psi^{\dl}_k}\right)\right)\right].\label{Eqn:LapTransInt}
\end{align}
Moreover, for a given $\|\widetilde{X}_*^{\dl}\|^2=y$, using Lemma \ref{Lem:LapTransShotNoise} and letting $\mathfrak{I}_0=I^{\dl}_{\mathcal{U}}$ yields
\begin{align}
\mathbb{E}\left[\exp\left(-\frac{s}{P_m}y^{\frac{\alpha}{2}}I^{\dl}_{\mathcal{U}}\right)\right]=\mathcal{L}_{\mathfrak{I}_0}\left(\frac{sy^{\frac{\alpha}{2}}}{P_m}\right)&=\exp\left\{-\pi\left(\sum_{k=1}^{M}\rho^{\ul}_k\lambda_k\right)\mathbb{E}\left[\left(\frac{sQDG}{P_m\mathsf{E}}\right)^{\frac{2}{\alpha}}y\right]\right\}\nonumber\\
&=\exp\left\{-\pi\nu\left(\sum_{k=1}^{M}\rho^{\ul}_k\lambda_k\right)\Xi_0\left(\mathsf{E},0,\frac{sQG}{P_m}\right)y\right\}.\label{Eqn:LapTransIntFD}
\end{align}
Therefore, using \eqref{Eqn:LapTransInt} and \eqref{Eqn:LapTransIntFD} we can have 
\begin{align*}
\mathcal{L}_{\frac{H_m}{\psi^{\dl}_m\gamma_0}}(s)=\int_{0}^{\infty}\exp\left(-\frac{s\epsilon_0Q}{P_m}y^{\frac{\alpha}{2}}-\pi y\widetilde{\lambda}^{\dl}\left[\widetilde{\Xi}^{\dl}_m(s)+1\right]\right) f_{\|\widetilde{X}^{\dl}_*\|^2}(y)\dif y,
\end{align*}
which leads to the lower bound on $C^{\dl}_{\nu, FD}$ given in \eqref{Eqn:DLrate}. 

For the uplink rate, it can be rewritten as
\begin{align*}
C^{\ul}_{\nu, FD}=\mathbb{E}\left[\log(1+\gamma_*^{\ul})\right] = \sum_{m=1}^{M} \vartheta^{\ul}_m \mathbb{E}\left[\int_{0}^{\infty}\frac{1}{s}\left(1-e^{-s\frac{H_m}{\psi^{\ul}_m}}\right)\mathcal{L}_{\frac{H_m}{\psi^{\ul}_m\gamma^{\ul}_*}}(s)\right].
\end{align*}
By following the similar derivation processes above,  $\mathcal{L}_{\frac{H_m}{\psi^{\ul}_m\gamma^{\ul}_*}}(s)$ is tightly lower-bounded by
\begin{align*}
\mathcal{L}_{\frac{H_m}{\psi^{\ul}_m\gamma^{\ul}_*}}(s)&=\mathbb{E}\left[\exp\left(-\frac{sI_*\|X^{\ul}_*\|^{\alpha}}{Q\psi^{\ul}_m}\right)\cdot\exp\left(-s\frac{\epsilon_* P_m}{Q\psi^{\ul}_m}\|X^{\ul}_*\|^{\alpha}\right)\right]\\
&\gtrapprox \int_{0}^{\infty} \mathcal{L}_{I_*}\left(\frac{sy^{\frac{\alpha}{2}}}{Q}\right)e^{-s\frac{\epsilon_*P_m}{Q}y^{\frac{\alpha}{2}}}f_{\|\widetilde{X}^{\ul}_*\|^2}(y) \dif y= \int_{0}^{\infty} e^{-\pi y\widetilde{\Xi}^{\ul}_m(s)}e^{-s\frac{\epsilon_*P_m}{Q}y^{\frac{\alpha}{2}}}f_{\|\widetilde{X}^{\ul}_*\|^2}(y) \dif y,
\end{align*}
in which $\mathcal{L}_{I_*}\left(\frac{sy^{\frac{\alpha}{2}}}{Q}\right)=\exp\left(-\pi y\widetilde{\lambda}^{\ul}\widetilde{\Xi}^{\ul}_m(s)\right)$.  Therefore, we finally have
\begin{align*}
C^{\ul}_{\nu, FD}\gtrapprox  \sum_{m=1}^{M} \vartheta^{\ul}_m \int_{0}^{\infty}\frac{1}{s}\left(1-\mathcal{L}_{\frac{H_m}{\psi^{\ul}_m}}(s)\right)e^{-s\frac{\epsilon_*P_m}{Q}y^{\frac{\alpha}{2}}-\pi y\widetilde{\lambda}^{\ul}\widetilde{\Xi}^{\ul}_m(s)}f_{\|\widetilde{X}^{\ul}_*\|^2}(y) \dif y,
\end{align*}
which is exactly the result in \eqref{Eqn:ULrate} by considering $\|X^{\ul}_*\|^2\sim\exp(\pi\widetilde{\lambda}^{\ul})$. This completes the proof. 

\subsection{Proof of Proposition \ref{Prop:rateRegion}}\label{App:ProofrateRegion}
As shown in Fig. \ref{Fig:rateregion} (b), regions $\mathcal{R}_{HD}$ and $\mathcal{R}_{FD}$ are enclosed by  $\mathcal{R}_{\inf}\defn\mathcal{R}^{\ul}_{\inf}\cup\mathcal{R}^{\dl}_{\inf}$ where $\mathcal{R}^{\ul}_{\inf}$ and $\mathcal{R}^{\dl}_{\inf}$ are defined by
\begin{align}
\mathcal{R}^{\ul}_{\inf}&\defn\{(R^{\ul},R^{\dl})\in\mathbb{R}^2: R^{\dl}\leq \nu C^{\dl}_{1,FD}, R^{\ul}\leq \nu C^{\ul}_{1,FD}+(1-\nu)C^{\ul}_{HD}\},\\
\mathcal{R}^{\dl}_{\inf}&\defn\{(R^{\ul},R^{\dl})\in\mathbb{R}^2: R^{\dl}\leq \nu C^{\dl}_{1,FD}+(1-\nu) C^{\dl}_{HD}, R^{\ul}\leq \nu C^{\ul}_{1,FD}\}.
\end{align}
Namely, $\mathcal{R}^{\ul}_{\inf}$ is the quadrilateral with vertices \textbf{0}, $(0,C^{\dl}_{HD})$, \textbf{f} and $(C^{\ul}_{1,FD},0)$ and $\mathcal{R}^{\dl}_{\inf}$ is the quadrilateral with vertices \textbf{0}, $(0,C^{\dl}_{1,FD})$, \textbf{f} and $(C^{\ul}_{HD},0)$. Note $\mathcal{R}^{\ul}_{\inf}\cap \mathcal{R}^{\dl}_{\inf}=\mathcal{R}_{FD}$ that is achievable by using the FD model all the time in the HetNet. All points on Line $(0,C^{\dl}_{HD})$-\textbf{f} can be achieved by changing time-sharing parameter $\nu$ (i.e. doing time-sharing) between 0 and 1. Thus, all points in $\mathcal{R}^{\dl}_{\inf}$ are achievable. By similar reasoning, all points on Line \textbf{f}-$(C^{\ul}_{HD},0)$ can also be achieved by doing time-sharing so that the entire region of  $\mathcal{R}^{\ul}_{\inf}$ is achievable. In addition, the four inequalities in $\mathcal{R}^{\ul}_{\inf}$ and $\mathcal{R}^{\dl}_{\inf}$ imply the following:
\begin{align*}
\frac{R^{\ul}-C^{\ul}_{1,FD}}{C^{\ul}_{HD}-C^{\ul}_{1,FD}}\leq (1-\nu),\,\, \frac{R^{\dl}}{C^{\dl}_{1,FD}}\leq \nu \Rightarrow \frac{R^{\ul}-C^{\ul}_{1,FD}}{C^{\ul}_{HD}-C^{\ul}_{1,FD}}+\frac{R^{\dl}}{C^{\dl}_{1,FD}}\leq 1\\
\frac{R^{\dl}-C^{\dl}_{1,FD}}{C^{\dl}_{HD}-C^{\dl}_{1,FD}}\leq (1-\nu),\,\, \frac{R^{\ul}}{C^{\ul}_{1,FD}}\leq \nu\Rightarrow \frac{R^{\dl}-C^{\dl}_{1,FD}}{C^{\dl}_{HD}-C^{\dl}_{1,FD}}+\frac{R^{\ul}}{C^{\ul}_{1,FD}}\leq 1.
\end{align*}
These two inequality constraints correspond to regions $\mathcal{R}^{\ul}_{\inf}$ and $\mathcal{R}^{\dl}_{\inf}$, respectively. Accordingly, $\mathcal{R}_{\inf}$ is the achievable convex hull of $\mathcal{R}_{FD}$ and $\mathcal{R}_{HD}$. 

Next, we want to show that the region enclosed by Line \textbf{0}-$(0,C^{\dl}_{HD})$, Red Dash Line $(0,C^{\dl}_{HD})$-$(C^{\ul}_{HD},0)$ and Line \textbf{0}-$(C^{\ul}_{HD},0)$, denoted by $\mathcal{R}_{\sup}$, is also achievable. Note that all points below Red Dash Line  $(0,C^{\dl}_{HD})$-\textbf{f} satisfy the following constraint
\begin{align*}
\frac{R^{\ul}-C^{\ul}_{\nu,FD}}{C^{\ul}_{HD}-C^{\ul}_{\nu,FD}}+\frac{R^{\dl}}{C^{\dl}_{\nu,FD}}\leq 1,
\end{align*}
where $C^{\ul}_{\nu,FD}$ and $C^{\dl}_{\nu,FD}$ contain the time-sharing variable $\nu$. According to \eqref{Eqn:DLrate} and \eqref{Eqn:ULrate}, it is easy to show that
\begin{align*}
\frac{\dif C^{\dl}_{\nu,FD}}{\dif\nu}\bigg|_{\nu=0} = 0\,\,\text{  and  }\,\, \frac{\dif^2 C^{\dl}_{\nu,FD}}{\dif\nu^2}\bigg|_{\nu\in(0,1]}<0.
\end{align*} 
This manifests that Red Dash Line  $(0,C^{\dl}_{HD})$-\textbf{f} is concave and above Line $(0,C^{\dl}_{HD})$-\textbf{f}. Similarly, we also can show that Red Dash Line  $(0,C^{\dl}_{HD})$-\textbf{f} is concave and above Line \textbf{f}-$(0,C^{\ul}_{HD})$. Therefore, region $\mathcal{R}_{\sup}$ encloses region $\mathcal{R}_{\inf}$ and it is the maximum achievable rate region of a user in an FD HetNet. 

\subsection{Proof of Proposition \ref{Prop:QueueingStability}}\label{App:ProofQueueingStability}
Due to the limited space, here we only show the stability of Algorithm \ref{Alg:Downlink} since the method of showing the stability of Algorithm \ref{Alg:Uplink} is similar. Consider the time right after the $n$th transmission and let $\mathbf{Q}_m(n)\defn[q_m^{\dl}(n)\,\,q_m^{\ul}(n)]^{\mathrm{T}}$ denote the queue length vector of the downlink queue $q^{\dl}_m(n)$ and the uplink queue $q^{\ul}_m(n)$ for a tier-$m$ BS. Note that the number of packets arriving at the tier-$m$ BS during a transmission time slot $\Delta t$ is a Poisson process with parameter $\eta^{\dl} \Delta t $ whereas the number of packets arriving at the user associated with the tier-$m$ BS during a transmission time slot $\Delta t$ is a Poisson process with parameter $\eta^{\ul} \Delta t $. Also, $\mathbf{Q}_m(n)$ forms a non-reducible Markov chain for all $m\in\mathcal{M}$. Without loss of generality, we assume all packets have the same size of $\ell=1$ in the following analysis. 

To show the stability of $\mathbf{Q}_m(n)$, we define the following Lyapunov function:
\begin{align}\label{Eqn:DefnLyapunov}
V_m(n) \defn v_m^{\dl}\left[q^{\dl}_m(n)\right]^2 + v_m^{\ul}\left[q^{\ul}_m(n)\right]^2 +2q^{\dl}_m(n) q^{\ul}_m(n) ,
\end{align}
where $v_m^{\dl}\defn \frac{C^{\dl}_{\nu,FD}}{C^{\dl}_{HD}-C_{\nu,FD}}$ and $v_m^{\ul}\defn \frac{C^{\ul}_{\nu,FD}}{C^{\ul}_{HD}-C^{\ul}_{\nu,FD}}$. Then we need to consider the following three cases:
\begin{enumerate}
	\item The downlink and uplink queues both are not empty ($q_m^{\dl}(n)>0$ and $q_m^{\ul}(n)>0$): For this case, the FD mode is adopted by the tier-$m$ and its user so that both queues are scheduled to be transmitted. Thus, we have $q_m^{\dl}(n+1)=q^{\dl}_m(n)-1+\Delta q^{\ul}_m(n)$ and $q_m^{\ul}(n+1)=q^{\ul}_m(n)-1+\Delta q^{\ul}_m(n)$ where $\Delta q_m^{\dl}(n) $ and $\Delta q^{\ul}_m(n)$ are Poisson random variables with parameters $\eta^{\dl}/C^{\dl}_{\nu,FD}$ and $\eta^{\ul}/C^{\ul}_{\nu,FD}$, respectively. Hence, we further have
	\begin{align}
	\mathbb{E}\left[V_m(n+1)|\mathbf{Q}_m(n)\right] =& V_m(n)+2\frac{C^{\dl}_{HD}}{C^{\dl}_{\nu,FD}}\left[v^{\dl}_m\left(\frac{R^{\dl}}{C^{\dl}_{HD}}-1\right)+\frac{R^{\ul}}{C^{\dl}_{HD}}\right]q^{\dl}_m(n)\nonumber\\
	&+2\frac{C^{\ul}_{HD}}{C^{\ul}_{\nu,FD}}\left[v^{\ul}_m\left(\frac{R^{\ul}}{C^{\ul}_{HD}}-1\right)+\frac{R^{\dl}}{C^{\ul}_{HD}}\right]q^{\ul}_m(n)+\mathsf{C}_1,
	\end{align}
	where $\mathsf{C}_1$ is a constant consisting of $R^{\ul}$, $R^{\dl}$, $C^{\dl}_{HD}$, $C^{\dl}_{\nu,FD}$, $C^{\ul}_{\nu,FD}$ and $C^{\ul}_{HD}$.
	\item The uplink queue is empty and the downlink queue is not empty ($q^{\dl}_m(n)>0$ and $q^{\ul}_m(n)=0$): For this case, the tier-$m$ BS just needs to use HD to transmit its packet since there is no uplink traffic and thus we have $q^{\dl}_m(n+1)=q_m^{\dl}(n)-1+\Delta q^{\dl}_m(n)$ and $q^{\ul}_m(n)=0$. This follows that
	\begin{align}
	\mathbb{E}\left[V_m(n+1)|\mathbf{Q}_m(n)\right]=V_m(n)+2\frac{C^{\dl}_{HD}}{C^{\dl}_{\nu,FD}}\left[v^{\dl}_m\left(\frac{R^{\dl}}{C^{\dl}_{HD}}-1\right)+\frac{R^{\ul}}{C^{\dl}_{HD}}\right]q^{\dl}_m(n)+\mathsf{C}_2,
	\end{align}
	where $\mathsf{C}_2$ is a constant consisting of $R^{\ul}$, $R^{\dl}$, $C^{\dl}_{HD}$, $C^{\dl}_{\nu,FD}$, $C^{\ul}_{\nu,FD}$ and $C^{\ul}_{HD}$.
	\item  The downlink queue is empty and the uplink queue is not empty ($q^{\ul}_m(n)>0$ and $q^{\dl}_m(n)=0$): For this case, we certainly have
	\begin{align}
	\mathbb{E}\left[V_m(n+1)|\mathbf{Q}_m(n)\right]=V_m(n)+\mathsf{C}_3,
	\end{align}
	where $\mathsf{C}_3$ is a constant consisting of $C^{\dl}_{HD}$, $C^{\dl}_{\nu,FD}$, $C^{\ul}_{\nu,FD}$ and $C^{\ul}_{HD}$.
Note that we know
\begin{align*}
v^{\dl}_m\left(\frac{R^{\dl}}{C^{\dl}_{HD}}-1\right)+\frac{R^{\ul}}{C^{\dl}_{HD}}<0\quad \text{and}\quad v^{\ul}_m\left(\frac{R^{\ul}}{C^{\ul}_{HD}}-1\right)+\frac{R^{\dl}}{C^{\ul}_{HD}}<0
\end{align*}
because all rate pairs $(R^{\ul},R^{\dl})$ satisfying these two inequalities are below the Red Dash Line between points $(C^{\ul}_{HD},0)$ and $(0,C^{\dl}_{HD})$ on Fig. \ref{Fig:rateregion} (b). Therefore, we can conclude $\mathbb{E}[V_m(n+1)|\mathbf{Q}_m(n)]<V_m(n)-1$ whenever $q_m^{\dl}(n)$ and $q^{\ul}_m(n)$ are large. According to the Foster-Lyapunov criterion \cite{SPMRLT93}, $\mathbf{Q}_m(n)$ is stable for all $m\in\mathcal{M}$.

\end{enumerate}


\bibliographystyle{ieeetran}
\bibliography{IEEEabrv,Ref_FullDuplexHetNet}

\end{document}